\documentclass[aps,pra,preprint]{revtex4}

\usepackage{epsfig}
\usepackage{subfigure}

\begin{document}

\title{Numerical Implementation of Non-Markovian Quantum State Diffusion}
\author{Joshua Wilkie and Ray Ng}
\affiliation{Department of Chemistry, Simon Fraser University, Burnaby, British Columbia V5A 1S6, Canada}

\date{\today}
 
\begin{abstract}

Non-Markovian quantum state diffusion (NMQSD) is a non-relativistic but otherwise exact 
theory which expresses 
the reduced density matrix of an arbitrary subsystem, interacting linearly with 
an uncoupled harmonic oscillator bath, as an average of diadics formed from state 
vectors which obey stochastic variational-differential equations. The vacuum 
radiation field can be represented as such an oscillator bath, and so this model 
is in widespread use in quantum optics. Prior to the development of NMQSD, exact 
subsystem solutions could only be obtained in a few special cases (e.g. spin-1/2, 
harmonic oscillator). Unfortunately, it has not yet been possible to obtain exact 
solutions to new problems using NMQSD due to the difficulty of solving the 
variational-differential equations. Here we show that these 
equations can be transformed into a pair of coupled nonlinear integrodifferential 
equations. We develop exact numerical methods for the integrodifferential equations 
and show that solutions can be readily obtained to good accuracy for quite 
general subsystems. We exactly solve various examples including tunneling in a double well
representing molecular isomerization or racemization, suppression of fluorescence from a two-level atom in a band gap, and intermittent fluorescence from a driven three level system 
representing electronic states of singly ionized magnesium.

\end{abstract}

\maketitle

\section{Introduction}

Few-level or few-body quantum systems such as single (or few chemically reacting) trapped
atomic\cite{Jump} or molecular ions in ion traps, interacting with lasers and 
the vacuum, are systems of considerable current interest. Intermittent single ion fluorescence has been
studied experimentally\cite{Jump} and theoretically\cite{Master,Master2,Knight,JumpTh}. Ions in traps have also served as 
models of quantum computers\cite{QC}. These systems can all be accurately modeled as a 
subsystem interacting linearly with a bath of harmonic oscillators. Exact solutions for 
this model have only been obtained for restrictive cases where the subsystem consists of 
a spin-1/2\cite{Hu,GPU} or a harmonic oscillator\cite{HPZ,HR}, and in a few other special 
cases. Theoretical studies of such systems have therefore usually relied on approximate 
master equations\cite{Master,Master2}, jump theories\cite{Knight} and Markovian stochastic 
Schr\"{o}dinger equations\cite{JumpTh}. Nevertheless, a theory yielding exact solutions 
for this model for general subsystem Hamiltonians would have widespread applications.
Such a theory could also prove useful as a limiting case for development of theories
for subsystem evolution under the influence of more general baths (e.g. single quantum-dot
\cite{Kuno} and single-molecule\cite{Silbey} fluorescence in condensed phase environments)
for which theory is still in the early stages of development\cite{Silbey,GenTh}, and for which 
intra-bath coupling is expected to play an important role\cite{Intrabath}. 

In a formal sense the problem of an arbitrary subsystem interacting linearly with an uncoupled
oscillator bath has recently been solved. The first important contribution toward this theory\cite{FV} showed that the influence functional has a particularly simple form. 
Subsequently it was shown that the exact reduced subsystem density, expressed as a path integral weighted by the influence functional, can be stochastically 
unraveled as an average over diadics constructed from state vectors which obey a 
non-Markovian variational-differential wave equation\cite{NMSD,NMSD2,NMSD3}. The resulting exact
theory has been called non-Markovian quantum state diffusion (NMQSD), and from NMQSD
one can {\em in principle} obtain exact solutions of the subsystem-oscillator bath
model for any subsystem. 

In practice, no new exact solutions have been obtained using 
NMQSD. While exact evolution equations can be formulated for any subsystem, these 
equations contain variational derivatives with respect to the complex colored noises which
emerge during the stochastic unraveling. Variational-differential equations (VDEs)
have not received the same attention as partial differential equations and much less is
known about their properties and solutions. As a consequence the evolution equation has
proved impossible to solve - even numerically - except for the few models for which exact 
solutions were previously known\cite{NMSD,NMSD2,NMSD3}. This impasse has led to the introduction of 
various perturbative expansions\cite{TYu,Gasp} and numerical approximation schemes\cite{SY}. However, direct
and exact methods for solution of the NMQSD equations remain an important goal.

Direct solution of the NMQSD equations may eventually be possible, but for the present methods which
circumvent the problem by eliminating the variational derivative seem most promising.
This manuscript introduces a pair of coupled nonlinear stochastic integrodifferential 
equations which we show are exactly equivalent to the stochastic VDE of NMQSD. Integrodifferential evolution equations can be converted to ordinary or partial differential equations\cite{Tann,IDE}. Hence, the pair of stochastic 
integrodifferential equations can be converted to stochastic ordinary or partial differential 
equations which are easier to solve. Recently developed numerical methods allow efficient solution of stochastic 
(ordinary and partial) differential equations (SDEs)\cite{SDE,COMM} even for quite large systems of equations\cite{TWC}. On the basis of the
transformed equations and using these SDE integrators we develop exact numerical methods 
for solving NMQSD problems. Example calculations for which exact solutions are known are used to 
verify the accuracy and efficiency of the new methods. We also numerically solve three new problems 
which were previously intractable.

Equivalent linear and nonlinear formulations of NMQSD exist. The nonlinear version of the 
theory preserves the norm of the state vector, which results in improved Monte Carlo convergence. Accordingly, we also consider linear and nonlinear 
reformulations of NMQSD. The simpler linear version is considered in section II, while the 
nonlinear case is treated in section III. Simple example calculations are used to illustrate the two 
approaches. More interesting applications to a tunneling problem representing molecular isomerization or recemization, to suppression of fluorescence from a two-level atom in a dielectric band gap, and to a driven three level system exhibiting intermittent fluorescence, are considered in section IV.

\section{Linear NMQSD Equations}

Consider for simplicity a subsystem-bath model with a single subsystem coupling operator $L$. The total Hamiltonian is
\begin{equation}
H_{tot}=H+\sum_{\omega}g_{\omega} (L a_{\omega}^{\dag}+L^{\dag}a_{\omega})+\sum_{\omega}\omega a_{\omega}^{\dag}a_{\omega}
\end{equation}
where we are using units in which $\hbar=1$. Here $H$ is the subsystem Hamiltonian, $g_{\omega}$ is a coupling
constant for an oscillator mode of frequency $\omega$, and the rest of the notation should be clear. The generalization of NMQSD and our results to multiple subsystem coupling operators is straightforward, so we will
confine our attention to this simplest case.

In NMQSD the evolution of the state vector $\psi_t$ is governed by the linear VDE
\begin{equation}
\frac{d\psi_t}{dt}=-iH ~\psi_t +z_t L ~\psi_t -L^{\dag}\int_0^tds~\alpha (t,s) ~\frac{\delta \psi_t}{\delta z_s}\label{EQ1}
\end{equation}
where $z_t$ is a complex colored noise with correlation function 
\begin{equation}
\alpha(t,s)=M[z_t^*z_s]. 
\label{MEMZ}
\end{equation}
In the limit of zero temperature $\alpha(t,s)=\sum_{\omega} g_{\omega}^2e^{-i\omega (t-s)}$ (see Refs. \cite{NMSD,NMSD2,NMSD3} for the non-zero temperature formula). Here $M[\dots ]$ 
denotes the average over different realizations of the noise. The exact reduced density matrix $\rho_t$ of the 
subsystem is given as an average of diadics via $\rho_t=M[|\psi_t\rangle \langle \psi_t|]$.

The solution $\psi_t$ of (\ref{EQ1}) is known to be an analytic functional of the noise $z_t$\cite{NMSD2} and so the 
variational derivative is well defined. The presence of
this variational derivative does however make the evolution equation difficult to solve. In some simple cases 
one can guess the form of $\frac{\delta \psi_t}{\delta z_s}$ and use a self-consistency requirement to find 
solutions of Eq. (\ref{EQ1}). Other than these few examples, it is generally unclear whether it is possible 
to directly solve the VDE (\ref{EQ1}) since numerical algorithms for VDEs have not yet been developed. Fortunately, it is possible to
reformulate the theory in terms of more easily solved integrodifferential equations.

To start with our reformulation, we introduce a non-unitary propagator $U(t,s)$ which evolves a state vector from time $s$ to time $t$.
If we consider subsystem evolution from a state $\psi_0$ at time 0 then $\psi_t=U(t,0)\psi_0$. From Eq. (\ref{EQ1}) we deduce that
\begin{equation}
\frac{d U(t,0)}{dt}=-iH ~U(t,0) +z_t L ~U(t,0) -L^{\dag}\int_0^tds~\alpha (t,s) ~\frac{\delta U(t,0)}{\delta z_s}.\label{EQ2}
\end{equation}
Next we need to eliminate the variational derivative $\frac{\delta U(t,0)}{\delta z_s}$. In particular, we will show that
\begin{equation}
\frac{\delta U(t,0)}{\delta z_s}=U(t,s)~L~U(s,0).\label{EQ3}
\end{equation}
Consider for simplicity the case where $L$ has a complete eigenbasis $|x\rangle$ which can be used to construct 
a path integral for the propagator along the lines followed in Ref. \cite{NMSD}. The path integral 
representation of $U(t,0)$ is a sum over paths weighted by an exponential whose argument includes a stochastic 
term $\int_0^tdu~z_u x_u$, where $L|x_u\rangle=x_u|x_u\rangle$ and $|x_u\rangle$ for each value of $u$ denotes an element of 
some complete basis (i.e. the path integral is over all $x_u$ for each value of $u$ between 0 and $t$)\cite{NMSD,NMSD2,NMSD3}. 
The variational derivative of $U(t,0)$ with respect to $z_s$ thus brings down a prefactor $\int_0^tdu~\frac{\delta z_u}{\delta z_s}x_u$ inside the path integral. Using the fact that 
\begin{eqnarray}
\frac{\delta z_u}{\delta z_s}=\delta(u-s)
\end{eqnarray}
and removing the closure relation for $x_s$ at time $s$ we then obtain Eq. (\ref{EQ3}). 

Now Eq. (\ref{EQ3}) and the semigroup property 
\begin{equation}
U(t,s)=U(t,0)U(s,0)^{-1}
\end{equation}
allow us to rewrite Eq. (\ref{EQ2}) in the form
\begin{equation}
\frac{d U(t,0)}{dt}=-iH ~U(t,0) +z_t L ~U(t,0) -L^{\dag}~U(t,0)\int_0^tds~\alpha (t,s) U(s,0)^{-1}L~U(s,0),\label{EQ4}
\end{equation}
which is a nonlinear integrodifferential equation for $U(t,0)$. This equation is exact and entirely 
equivalent to Eq. (\ref{EQ1}). No loss of generality is incurred in working with Eq. (\ref{EQ4}), but
in a finite basis set the number of equations arising from (\ref{EQ4}) is the square of the number of
equations arising from Eq. (\ref{EQ1}). Moreover, Eq. (\ref{EQ4}) is clearly nonlinear whereas Eq. (\ref{EQ1}) is linear in at least some cases. These apparent defects are unfortunate, but one must recognize that Eq. (\ref{EQ4}) is solvable while (\ref{EQ1}) is not, and the computational costs of solving (\ref{EQ4}) are quite reasonable for
a large class of potentially interesting problems. We solve three such examples in section IV. Thus,
implementations and applications of (\ref{EQ4}) and its norm-preserving generalizations are the focus of this manuscript.

Direct inversion of $U(s,0)$, required for (\ref{EQ4}), can be avoided at the expense of adding a second equation. Consider the dynamics 
of $U(0,t)=U(t,0)^{-1}$. Using the fact that $U(0,t)U(t,0)=1$ and differentiating with respect to $t$ gives
\begin{eqnarray}
&&\frac{d U(0,t)}{dt}U(t,0)+U(0,t)\{-iH U(t,0) +z_t L U(t,0) \nonumber \\
&&-L^{\dag}U(t,0)\int_0^tds~\alpha (t,s) U(s,0)^{-1}LU(s,0)\}=0.
\end{eqnarray}
Multiplying on the left by $U(t,0)^{-1}$ then gives
\begin{eqnarray}
\frac{d U(0,t)}{dt}+U(0,t)\{-iH+z_t L -L^{\dag}U(t,0)\int_0^tds~\alpha (t,s) U(s,0)^{-1}LU(s,0)U(0,t)\}=0
\end{eqnarray}
or 
\begin{eqnarray}
\frac{d U(0,t)}{dt}&=&iU(0,t)H-z_t U(0,t)L \nonumber \\
&+&U(0,t)L^{\dag}U(t,0)\int_0^tds~\alpha (t,s) U(s,0)^{-1}LU(s,0)U(0,t)\label{EQ5}
\end{eqnarray}
which is again of integrodifferential form. Note that both equations (\ref{EQ4}) and (\ref{EQ5}) involve only $U(t,0)$ and $U(t,0)^{-1}$ so that the pair is closed.

Changing notation to $U_t=U(t,0)$ and $U_t^{-1}=U(0,t)$ we can rewrite equations (\ref{EQ4}) and (\ref{EQ5}) as 
\begin{eqnarray}
\frac{d U_t}{dt}&=&-iH ~U_t +z_t L ~U_t -L^{\dag}~U_t\int_0^tds~\alpha (t,s) U_s^{-1}L~U_s\nonumber \\
\frac{d U_t^{-1}}{dt}&=&iU_t^{-1}~H-z_t U_t^{-1}L +U_t^{-1}L^{\dag}~U_t\int_0^tds~\alpha (t,s) U_s^{-1}L~U_s~U_t^{-1} \label{UUI}
\end{eqnarray}
which is a closed set of integrodifferential equations. We will now show that Eqs. (\ref{UUI}) can be transformed into sets of ordinary or partial differential equations.

The most efficient set of transformed equations depends on the properties of the memory function. In section IIA we assume that the memory function consists of a few terms of exponential form, i.e.,
\begin{equation}
\alpha (t,s)=\sum_{j=1}^m A_j e^{-\gamma_j |t-s|} e^{-i\omega_j(t-s)}\label{MEM}
\end{equation}
where $A_j$ and $\gamma_j$ are positive numbers. The terms in Eq. (\ref{MEM}) do not in general correspond to
physical bath oscillator modes. Instead Eq. (\ref{MEM}) can be viewed as a best fit to the memory function, 
obtained by nonlinear least squares\cite{Tann} or other techniques\cite{Mand}. In many cases the number of required terms $m$ can be quite small. The general case where $m$ is very large, or where $\alpha (t,s)$ cannot be written in the form (\ref{MEM}), is considered in section IIB. To illustrate the application of these methods we numerically solve a number of example problems. For each example problem, Eqs. (\ref{UUI}) are expressed in ordinary differential form and solved using stochastic integration methods\cite{SDE,COMM}. 

Colored noises can be generated using a variety of techniques\cite{Gasp,Gasp2,CNOISE}. We chose memory functions for our examples which can be expressed via Eq. (\ref{MEM}). The complex colored noise is then generated via $z_t=\sum_{j=1}^m\xi_t^j$ by integrating the stochastic differential equations
\begin{equation}
d\xi_t^j=-(\gamma_j+i\omega_j)\xi_t^j dt +\sqrt{2\gamma_jA_j}dW_t^j\label{SPR}
\end{equation}
from $-\infty$ (in practice some large negative time) to time $t=0$, and then from $t=0$ onward in combination with transformed versions of Eqs. (\ref{UUI}). Here $W_t^j$ are complex Wiener processes satisfying $M[dW_t^{j*}dW_s^k]=\sqrt{dt}\delta_{j,k}\delta_{t,s}$ and $M[dW_t^{j}dW_s^k]=0$. One can show that 
\begin{equation}
\xi_t^j=\sqrt{2\gamma_jA_j}\int_{-\infty}^tdW_s^j ~e^{-\gamma_j(t-s)}e^{-i\omega_j(t-s)},\label{xi}
\end{equation}
$z_t=\sum_{j=1}^m\xi_t^j$, and Eq. (\ref{MEMZ}) yield the correct memory function (\ref{MEM}) in the mean. The stochastic differential equations (\ref{SPR}) are obtained by differentiating expressions (\ref{xi}) for $\xi_t^j$. Sets of stochastic differential equations like (\ref{SPR}) and the transformed versions of (\ref{UUI}) can be solved to any required tolerance using recently developed high order variable stepsize integration methods\cite{SDE,COMM}. 

\subsection{Sum of exponentials}

In many cases it is possible to expand the memory function as a sum of a few exponentials of the form Eq. (\ref{MEM}). Fits to expansions of this type can be obtained using nonlinear least squares algorithms. In practice, obtaining good fits with nonlinear least squares can be a frustratingly time consuming endeavor. The nonlinearity of the fitting function makes minimization algorithms highly sensitive to the parameter search 
domains. In such cases, it may be possible to obtain expansions of the same form using other techniques\cite{Mand}.

Once this expansion is known we can define operators
\begin{equation}
V_{t,j}=\int_0^t ds~ A_j e^{-\gamma_j (t-s)} e^{-i\omega_j(t-s)} U_s^{-1}LU_s
\end{equation}
such that the full set of equations in ordinary differential form becomes
\begin{eqnarray}
\frac{d U_t}{dt}&=&-iH U_t +z_t L U_t -L^{\dag}U_t\sum_{j=1}^mV_{t,j}\nonumber \\
\frac{d U_t^{-1}}{dt}&=&iU_t^{-1}H-z_t U_t^{-1}L +U_t^{-1}L^{\dag}U_t\sum_{j=1}^mV_{t,j} U_t^{-1}\nonumber \\
\frac{d V_{t,j}}{dt}&=&-(\gamma_j+i\omega_j)V_{t,j}+A_jU_t^{-1}LU_t.\label{lindc}
\end{eqnarray}
Efficient implementation of these equations requires that $\sum_{j=1}^mV_{t,j}$ be computed first and stored temporarily. From this quantity $U_t\sum_{j=1}^mV_{t,j}$ and 
$L^{\dag}U_t\sum_{j=1}^mV_{t,j}$ can be calculated. Numerically it is easier to use the fact that $U_tU_t^{-1}=1$ and hence that $dU_t^{-1}/dt=-U_t^{-1} dU_t/dt U_t^{-1}$ to find $dU_t^{-1}/dt$ than to program the above equation. 

To illustrate the utility of the method we will apply the transformed equations to a few problems where exact solutions are known. In all cases the SDEs were solved using the ANISE software\cite{COMM}. We also used a single processor for each calculation, but of course one of the primary numerical advantages of quantum state diffusion theories is that they are inherently parallel and should ideally be implemented on clusters. With the exception of example IIC the calculations took only a few minutes.

\subsubsection{Example IIA}

Consider the special case where $H=(\omega/2)\sigma_z$ and $L=\lambda \sigma_z$ and pick $\alpha(t,s)=(\gamma/2)e^{-\gamma(t-s)}$ with $\omega=1$, $\lambda^2=2\omega$ and $\gamma=\omega$. We chose the initial condition $|\psi_0\rangle =\frac{1+2i}{\sqrt{7}}|1\rangle +\frac{1+i}{\sqrt{7}}|2\rangle$ where $\sigma_z|i\rangle =(-1)^{i-1}|i\rangle$ for $i=1,2$ (this is the same convention as in Ref. \cite{NMSD2}) . We calculated $\langle 1|\rho_t|2\rangle$ using Eqs. (\ref{lindc}) and an average over 10000 
trajectories. In Fig. 1 we plot the real (solid curve) and imaginary (dotted curve) parts vs time against the corresponding known exact results\cite{NMSD2} (dashed and dot-dashed, respectively).
\begin{figure}[htp]
\caption{$\langle 1|\rho_t|2\rangle$ for Example IIA}
\epsfig{file=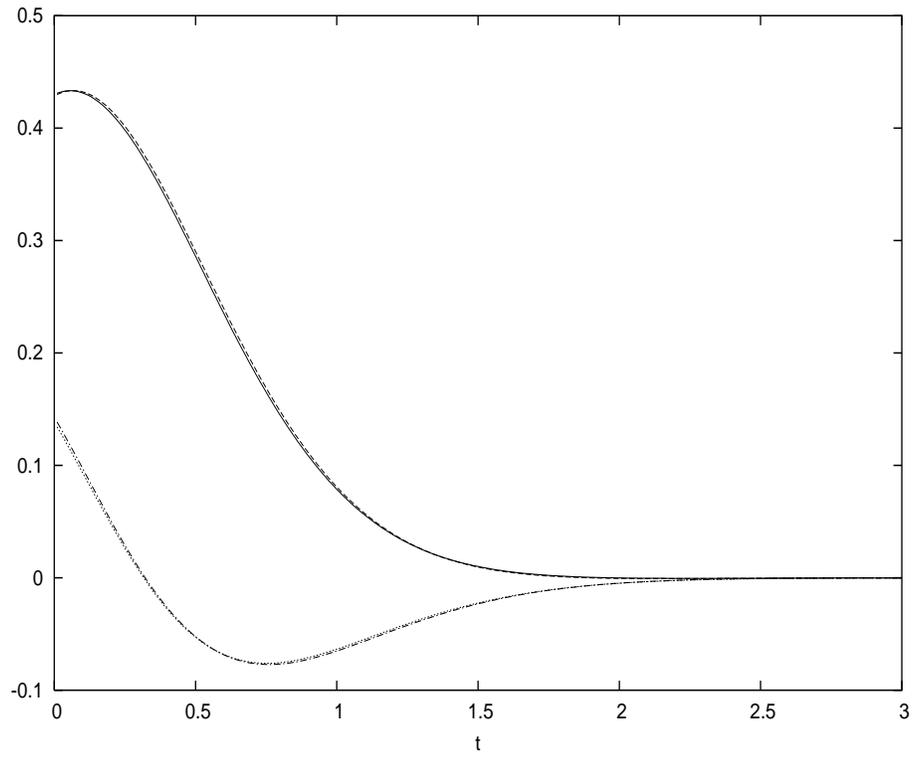,width=5in,height=4in}
\end{figure}
The accuracy is already very good. In Fig. 2 we plot the diagonal elements of $\rho_t$ which are supposed to be constant for this model. We see that the numerically calculated $\langle 1|\rho_t|1\rangle$ (solid curve) while initially equal to 5/7 (dashed line) decays away from this value as time proceeds. Similar results are observed for the element $\langle 2|\rho_t|2\rangle$ (dotted curve) which starts at 2/7 (dot-dashed line) and grows. Hence, for 10000 trajectories convergence is still incomplete for the diagonal elements. We will show in section III that introducing norm-preserving equations will lead to faster convergence.
\begin{figure}[htp]
\caption{$\langle 1|\rho_t|1\rangle$ and $\langle 2|\rho_t|2\rangle$ for Example IIA}
\epsfig{file=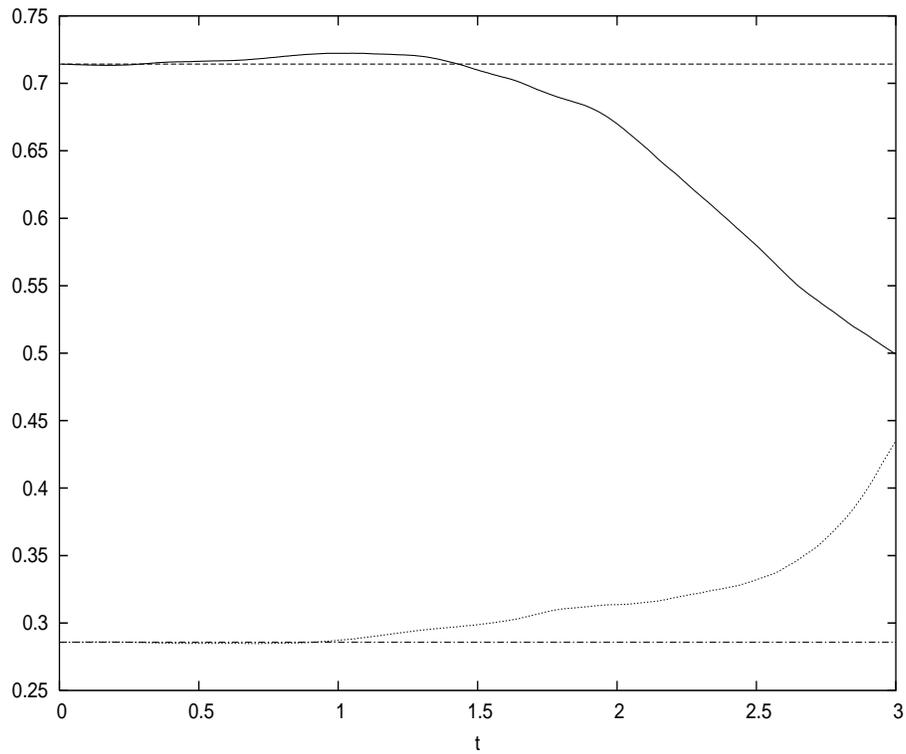,width=5in,height=4in}
\end{figure}

We also show in Fig. 3 the memory function calculated numerically using the stochastic process $z_t$ (solid curve) and the exact memory function $.5e^{-t}$ (dashed curve). Agreement is satisfactory.
\begin{figure}[htp]
\caption{$\alpha(t,0)$ for Example IIA}
\epsfig{file=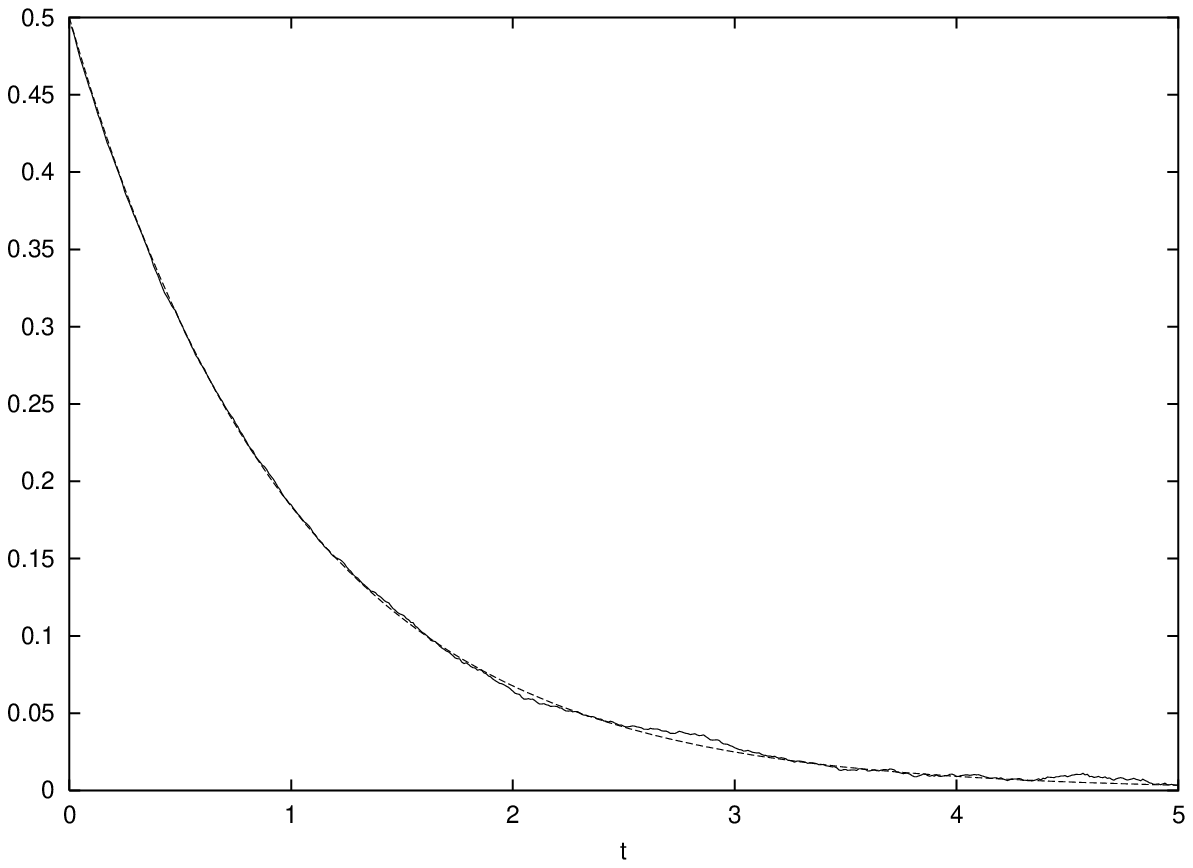,width=5in,height=4in}
\end{figure}

\subsubsection{Example IIB}

Now consider the case where $H=(\omega/2)\sigma_z$ and $L=(\lambda/2)(\sigma_x-i\sigma_y)$ (Example IIIB in Ref. \cite{NMSD2}) and again pick $\alpha(t,s)=(\gamma/2)e^{-\gamma(t-s)}$ with $\omega=1$, $\lambda^2=2\omega$ and $\gamma=\omega$. For the initial condition $|\psi_0\rangle =\frac{1}{\sqrt{2}}(|1\rangle +|2\rangle)$ we again calculated $\langle 1|\rho_t|2\rangle$ using Eqs. (\ref{lindc}) and an average over 10000 
trajectories. In Fig. 4 we plot the real (solid curve) and imaginary (dotted curve) parts vs time against the corresponding known exact results\cite{NMSD2} (dashed and dot-dashed, respectively).
Good agreement is obtained. 
\begin{figure}[htp]
\caption{$\langle 1|\rho_t|2\rangle$ for Example IIB}
\epsfig{file=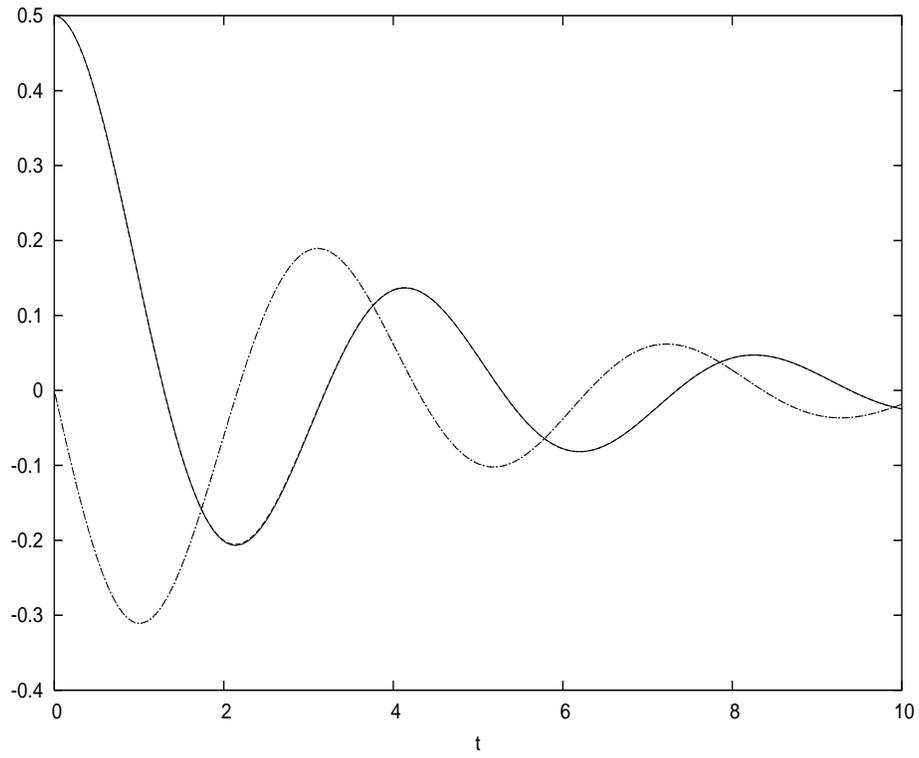,width=5in,height=4in}
\end{figure}
Figure 5 shows the diagonal matrix elements $\langle 1|\rho_t|1\rangle$ and $\langle 2|\rho_t|2\rangle$ plotted against their corresponding exact results. The results here are also good.
\begin{figure}[htp]
\caption{$\langle 1|\rho_t|1\rangle$ and $\langle 2|\rho_t|2\rangle$ for Example IIB}
\epsfig{file=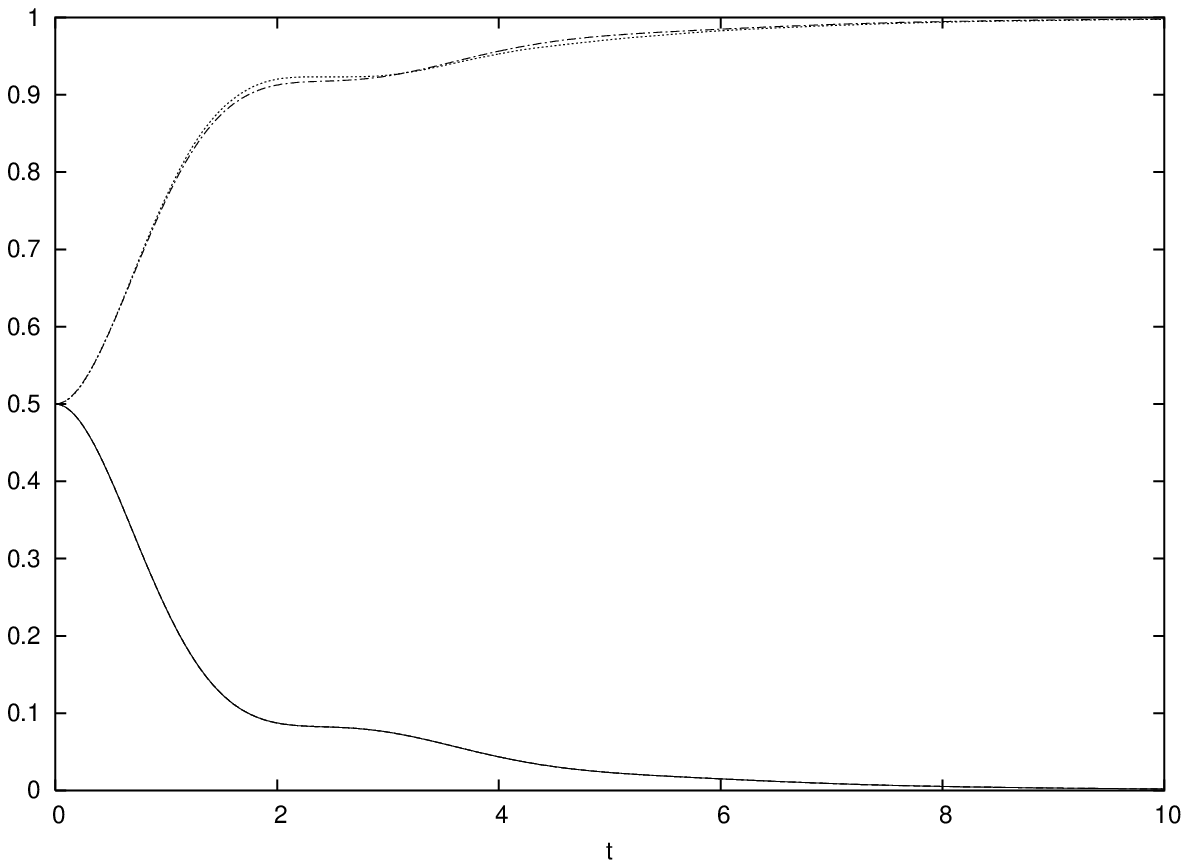,width=5in,height=4in}
\end{figure}

\subsubsection{Example IIC}

Finally, consider a harmonic oscillator subsystem with $H=\omega a^{\dag}a$ and $L=\lambda a$. We chose an initial condition $|\psi_0\rangle =\frac{1+2i}{\sqrt{7}}|0\rangle +\frac{1+i}{\sqrt{7}}|1\rangle $ (where $a^{\dag}a|n\rangle=n|n\rangle$ for $n=0,1,\dots$) and picked $\alpha(t,s)=(\gamma/2)e^{-\gamma(t-s)}$ with $\omega=1$, $\lambda=\omega$ and $\gamma=\omega$. A basis set consisting of the first five oscillator states was employed. We computed 10000 trajectories to construct the average. The cpu time was about 114 minutes on a 600 MHz Alpha processor. In Fig. 6 we plot the real and imaginary parts of $\langle 1|\rho_t|2\rangle$ (solid and dotted) vs their corresponding exact real and imaginary parts (dashed and dot-dashed). In Fig. 7 we show that diagonal matrix elements $\langle 1|\rho_t|1\rangle$ and $\langle 2|\rho_t|2\rangle$ (solid and dotted) plotted against their corresponding exact results (dashed and dot-dashed). The results are uniformly good.
\begin{figure}[htp]
\caption{$\langle 1|\rho_t|2\rangle$ for Example IIC}
\epsfig{file=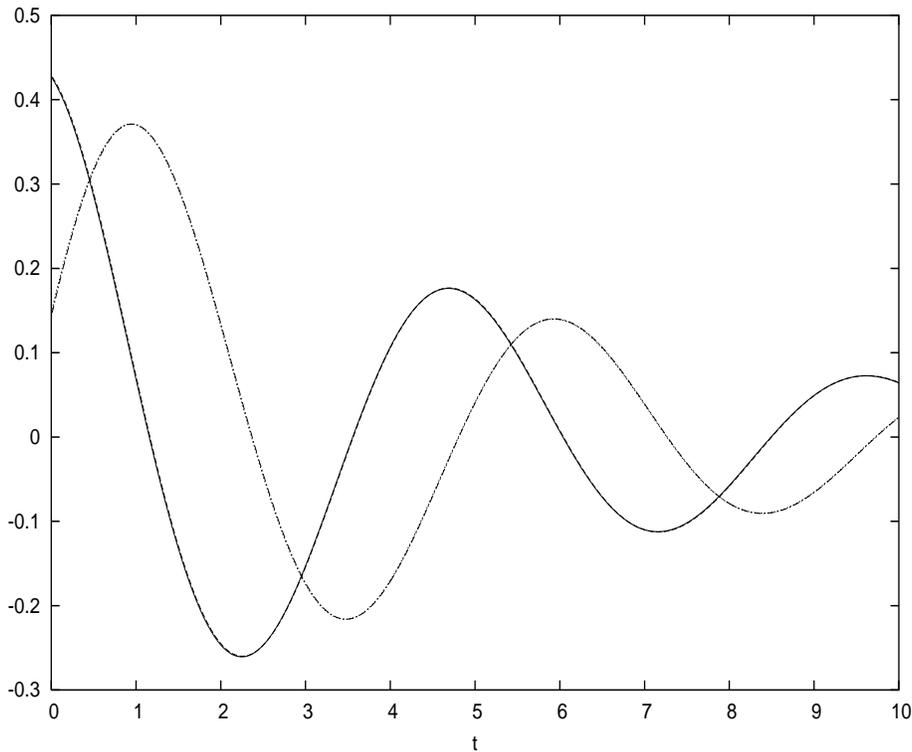,width=5in,height=4in}
\end{figure}
\begin{figure}[htp]
\caption{$\langle 1|\rho_t|1\rangle$ and $\langle 2|\rho_t|2\rangle$ for Example IIC}
\epsfig{file=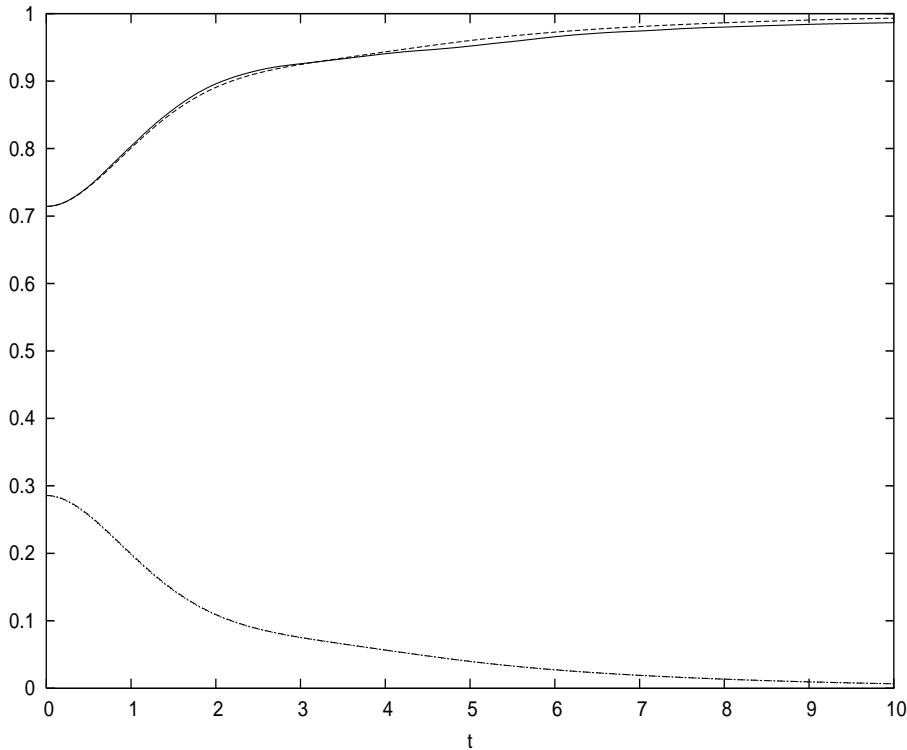,width=5in,height=4in}
\end{figure}

\subsection{General memory function}

Here we consider the opposite case where the memory function cannot be efficiently represented as
a sum of exponential terms like (\ref{MEM}). In this case it may still be desirable to 
generate the colored noise using Eqs. (\ref{SPR}), but otherwise the noise can be generated with
spectral methods and other techniques\cite{Gasp,Gasp2,CNOISE}. The key point is that Eqs. (\ref{lindc}) will
not be efficient and some new approach must be explored. 

Equations (\ref{UUI}) are of integrodifferential form. Such equations can be solved exactly using
a recently developed method\cite{IDE}. The trick is to convert the equations to partial differential
form. To do this we introduce a new variable $u$ (where $u\in (-\infty,\infty )$) and define
\begin{equation}
V_{t,u}=\int_0^t ds~\alpha(t+u,s) U_s^{-1}LU_s.
\end{equation}
Differentiating we find that the new set of equations has a partial differential form
\begin{eqnarray}
\frac{d U_t}{dt}&=&-iH U_t +z_t L U_t -L^{\dag}U_tV_{t,0}\nonumber \\
\frac{d U_t^{-1}}{dt}&=&iU_t^{-1}H-z_t U_t^{-1}L +U_t^{-1}L^{\dag}U_tV_{t,0}U_t^{-1}\nonumber \\
\frac{d V_{t,u}}{dt}&=&\alpha(t+u,t)U_t^{-1}LU_t+\frac{\partial V_{t,u}}{\partial u}\label{tu}
\end{eqnarray}
which is exactly equivalent to the original integrodifferential set (\ref{UUI}).
The partial derivative $\frac{\partial V_{t,u}}{\partial u}$ can be evaluated using fast Fourier transform
methods or using discrete variable representations\cite{IDE,DVR}, or by employing a harmonic oscillator basis. In the first two cases the variable $u$ is 
represented on a grid of points, while a discrete basis is used in the third, and so Eqs. (\ref{tu}) revert to ordinary differential form and hence they can also
be solved using the method of Ref. \cite{SDE} or using the ANISE software\cite{COMM}. 

In practice there are advantages to modifying the above equations somewhat by instead defining
\begin{equation}
V_{t,u}=f(u)\int_0^t ds~\alpha(t+u,s) U_s^{-1}LU_s
\end{equation}
where $f(u)=1$ and where $f(u)$ decays rapidly with $u$. These can lead to a smaller required basis 
set\cite{IDE} and hence faster algorithms. The modified $V_{t,u}$ obeys
\begin{eqnarray}
\frac{d V_{t,u}}{dt}=f(u)\alpha(t+u,t)U_t^{-1}LU_t+\frac{\partial V_{t,u}}{\partial u}-\frac{f'(u)}{f(u)}V_{t,u}\label{tup}
\end{eqnarray}
while the equations for $U_t$ and $U_t^{-1}$ are unaltered. We will not explore the issue of which is the optimal form for $f(u)$, but merely note that some success has been had with $f(u)=e^{-\alpha u^2}$\cite{IDE}. In the case where the $u$ degree of freedom is represented in an oscillator basis, $f(u)$ should be chosen so that the oscillator matrix elements $\langle n|f(u)\alpha(t+u,t)|m\rangle$ are nonzero only for the first few basis functions. 

Finally, we note that it is possible to use a similar trick to generate the complex noise $z_t$ when it is stationary, i.e. when $\alpha(t,s)=c(t-s)$\cite{Gasp,Gasp2}. Define 
\begin{equation}
z_t=\int_{-\infty}^tdW_s~R(t-s)\label{REP}
\end{equation}
with $R(t)=0$ for $t<0$, and where $dW_s$ is a differential Wiener process with properties $M[dW_t^{*}dW_s]=\sqrt{dt}\delta_{t,s}$ and $M[dW_tdW_s]=0$. It follows that
\begin{equation}
M[z_t^*z_s]=\int_{-\infty}^{\infty}d\tau~R^*(\tau)R(\tau-t)
\end{equation}
from which one can show that
\begin{eqnarray}
R(t)&=&\frac{1}{\sqrt{2\pi}}\int_{-\infty}^{\infty} d\omega~G(\omega)e^{i\omega t} \\
|G(\omega)|^2&=&\frac{1}{2\pi}\int_{-\infty}^{\infty} dt~c(t) e^{i\omega t}.
\end{eqnarray}
One can then choose $G(\omega)=\sqrt{|G(\omega)|^2}$ (see Refs. \cite{Gasp,Gasp2}).

Now that we have the representation (\ref{REP}) for the noise we can define
\begin{equation}
z_{t,u}=f(u)\int_{-\infty}^tdW_s~R(t+u-s)\label{REP2}
\end{equation}
with $f(0)=1$ such that $z_t=z_{t,0}$. Differentiating Eq. (\ref{REP2}) we then obtain the partial differential equation
\begin{equation}
dz_{t,u}=[\frac{\partial z_{t,u}}{\partial u}-\frac{f'(u)}{f(u)}z_{t,u}]dt+f(u)R(u)dW_t
\end{equation}
which can be solved in concert with Eqs. (\ref{tu}) and (\ref{tup}).

Obviously implementation of the rather sophisticated approach outlined in this section is quite a bit more involved than that for exponential type memory functions. However, all aspects of this treatment are exact and
similar calculations have been shown effective for other types of integrodiffential equations\cite{IDE}. Hence, we will leave a detailed study of this approach to a subsequent manuscript.

\section{Nonlinear NMQSD Equations}

While we have obtained mostly good convergence and accuracy using the linear NMQSD equation, the theory can also be formulated in terms of a norm preserving nonlinear VDE. The introduction of norm-preserving equations is important since their solutions may have 
physical significance\cite{Wise}. More practically, the norm-preserving equations in many instances yield faster convergence of the mean with the number of trajectories. The norms of the non-norm-preserving equations go to zero with probability one and so for long time dynamics, contributions to the mean tend to come from a few unusual trajectories potentially making convergence slow. The computational cost per trajectory for the norm-preserving equations is only a little greater than that of Eqs. (\ref{UUI}). Moreover, the non-norm-preserving equations are themselves nonlinear and so there is no apparent drawback to the norm-preserving reformulation of NMQSD.

The norm-preserving formulation of NMQSD is obtained via a two step process consisting of a Girsanov transformation followed by normalization of the wave vector. The details are given in Ref. \cite{NMSD2}. Defining $U_t$ through $\psi_t=U_t\psi_0$, and substituting into the norm-preserving wave equation of Ref. \cite{NMSD2}, it can be deduced that
\begin{eqnarray}
\frac{dU_t}{dt}&=&-iH U_t +(z_t+\int_0^tds~\alpha^*(t,s)\langle L^{\dag}\rangle_s) ~(L-\langle L\rangle_t) ~U_t \nonumber \\
&-&(L^{\dag}-\langle L^{\dag}\rangle_t)~U_t\int_0^tds~\alpha (t,s) U_s^{-1}LU_s\nonumber \\
&+&
\langle \psi_0|U_t^{\dag}(L^{\dag}-\langle L^{\dag}\rangle_t)U_t\int_0^tds~\alpha (t,s) U_s^{-1}LU_s|\psi_0\rangle ~U_t
\end{eqnarray}
where $\langle L\rangle_t=\langle \psi_t|L|\psi_t\rangle$. Note that $U_t$ now depends on the initial wave function $\psi_0$. As in the linear case there is also an equation for the inverse $U^{-1}_t$,
\begin{eqnarray}
\frac{d U_t^{-1}}{dt}&=&iU_t^{-1}H-(z_t +\int_0^tds~\alpha^*(t,s)\langle L^{\dag}\rangle_s)~U_t^{-1}~(L-\langle L\rangle_t) \nonumber \\
&+&U_t^{-1} (L^{\dag}-\langle L^{\dag}\rangle_t) ~U_t \int_0^tds~\alpha (t,s) U_s^{-1}LU_s  ~U_t^{-1}\nonumber \\
&-&\langle \psi_0|U_t^{\dag}(L^{\dag}-\langle L^{\dag}\rangle_t)U_t\int_0^tds~\alpha (t,s) U_s^{-1}LU_s|\psi_0\rangle ~U_t^{-1}.
\end{eqnarray}
These integrodifferential equations can again be re-expressed as sets of ordinary or partial differential equations. Again we consider two cases.

\subsection{Sum of exponentials}

If the memory function can be expressed as a sum of exponentials via (\ref{MEM}) then we can again define operators
\begin{equation}
V_{t,j}=\int_0^t ds~ A_j e^{-\gamma_j (t-s)} e^{-i\omega_j(t-s)} U_s^{-1}LU_s
\end{equation}
such that the full set of equations in ordinary differential form becomes
\begin{eqnarray}
\frac{dU_t}{dt}&=&-iH U_t +(z_t+\sum_{j=1}^m y_{t,j}) ~(L-\langle L\rangle_t) ~U_t \nonumber \\
&-&(L^{\dag}-\langle L^{\dag}\rangle_t)~U_t\sum_{j=1}^mV_{t,j} \nonumber \\
&+&
\langle \psi_0|U_t^{\dag}(L^{\dag}-\langle L^{\dag}\rangle_t)U_t\sum_{j=1}^mV_{t,j} |\psi_0\rangle ~U_t\nonumber \\
\frac{d U_t^{-1}}{dt}&=&iU_t^{-1}H-(z_t +\sum_{j=1}^m y_{t,j})~U_t^{-1}~(L-\langle L\rangle_t) \nonumber \\
&+&U_t^{-1} (L^{\dag}-\langle L^{\dag}\rangle_t) ~U_t ~\sum_{j=1}^mV_{t,j}  ~U_t^{-1}\nonumber \\
&-&\langle \psi_0|U_t^{\dag}(L^{\dag}-\langle L^{\dag}\rangle_t)U_t\sum_{j=1}^mV_{t,j}  |\psi_0\rangle ~U_t^{-1}\nonumber \\
\frac{d V_{t,j}}{dt}&=&-(\gamma_j+i\omega_j)V_{t,j}+A_jU_t^{-1}LU_t\nonumber \\
\frac{dy_{t,j}}{dt}&=&-(\gamma_j-i\omega_j)y_{t,j}+A_j\langle L^{\dag}\rangle_t\label{EEs}
\end{eqnarray}
where $y_{t,j}=\int_0^tds A_je^{-\gamma_j (t-s)} e^{i\omega_j(t-s)} \langle L^{\dag}\rangle_s$. A key to efficient numerical implementation of these equations is evaluation and temporary storage of $\sum_{j=1}^mV_{t,j}$ from which  $U_t\sum_{j=1}^mV_{t,j}$, and $(L^{\dag}-\langle L^{\dag}\rangle_t)~U_t\sum_{j=1}^mV_{t,j}$, $\langle \psi_0|U_t^{\dag}(L^{\dag}-\langle L^{\dag}\rangle_t)U_t\sum_{j=1}^mV_{t,j} |\psi_0\rangle$ can be evaluated. Again $dU_t^{-1}/dt=-U_t^{-1} dU_t/dt U_t^{-1}$ should be used to find $dU_t^{-1}/dt$. The conservation law $\langle \psi_t|\psi_t\rangle=1$ proves very useful for debugging code for Eqs. (\ref{EEs}). When ${\rm Tr}\{L\}=0$ an additional check can be made to verify that ${\rm Tr}\{V_{t,j}\}=0$.

\subsubsection{Example IIIA}

Consider again the case where $H=(\omega/2)\sigma_z$ and $L=\lambda \sigma_z$ and pick $\alpha(t,s)=(\gamma/2)e^{-\gamma(t-s)}$ with $\omega=1$, $\lambda^2=2\omega$ and $\gamma=\omega$. For the initial condition $|\psi_0\rangle =\frac{1+2i}{\sqrt{7}}|1\rangle +\frac{1+i}{\sqrt{7}}|2\rangle $ we calculated $\langle 1|\rho_t|2\rangle$ using Eqs. (\ref{lindc}) and an average over 10000 
trajectories. In Fig. 8 we plot the real (solid curve) and imaginary (dotted curve) parts vs time against the corresponding known exact results\cite{NMSD2} (dashed and dot-dashed, respectively).
\begin{figure}[htp]
\caption{$\langle 1|\rho_t|2\rangle$ for Example IIIA}
\epsfig{file=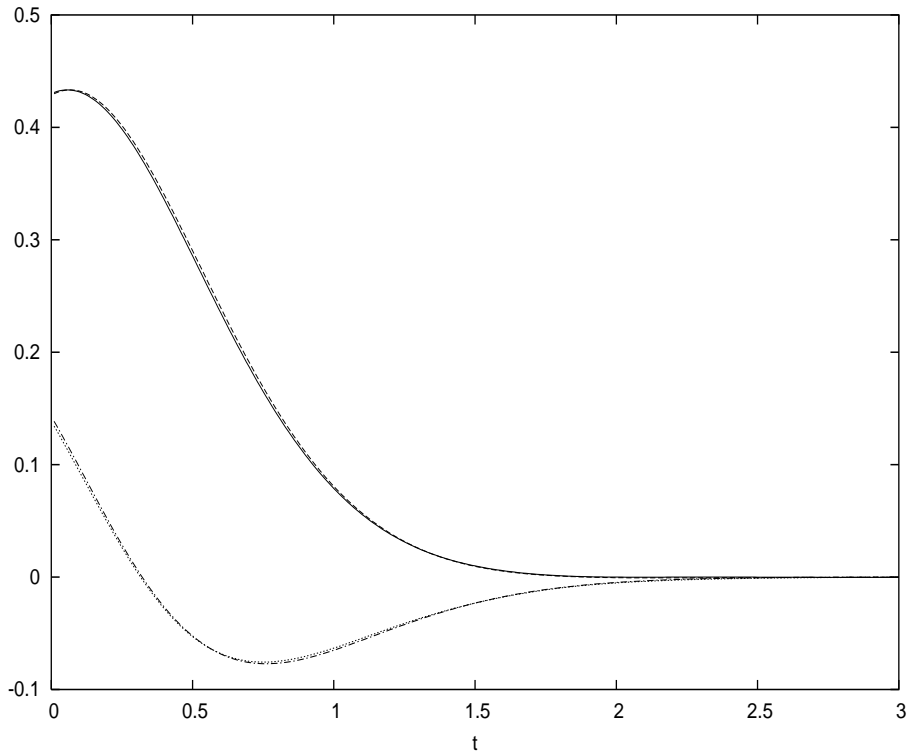,width=5in,height=4in}
\end{figure}
Once again we used the ANISE software\cite{COMM} to solve the equations. Norm was preserved to machine precision for individual trajectories. 

As in the case of the non-norm-preserving equations, convergence of the off-diagonal matrix element is good with 10000 trajectories. In Fig. 9 we show the diagonal matrix elements $\langle 1|\rho_t|1\rangle$ and $\langle 2|\rho_t|2\rangle$. For the non-norm-preserving equations we obtained poor convergence for 10000 trajectories. However, for the norm-preserving equations the convergence is quite good. The non-preserving equations do indeed seem to lead to faster convergence as claimed in Ref. \cite{NMSD2}.
\begin{figure}[htp]
\caption{$\langle 1|\rho_t|1\rangle$ and $\langle 2|\rho_t|2\rangle$ for Example IIIA}
\epsfig{file=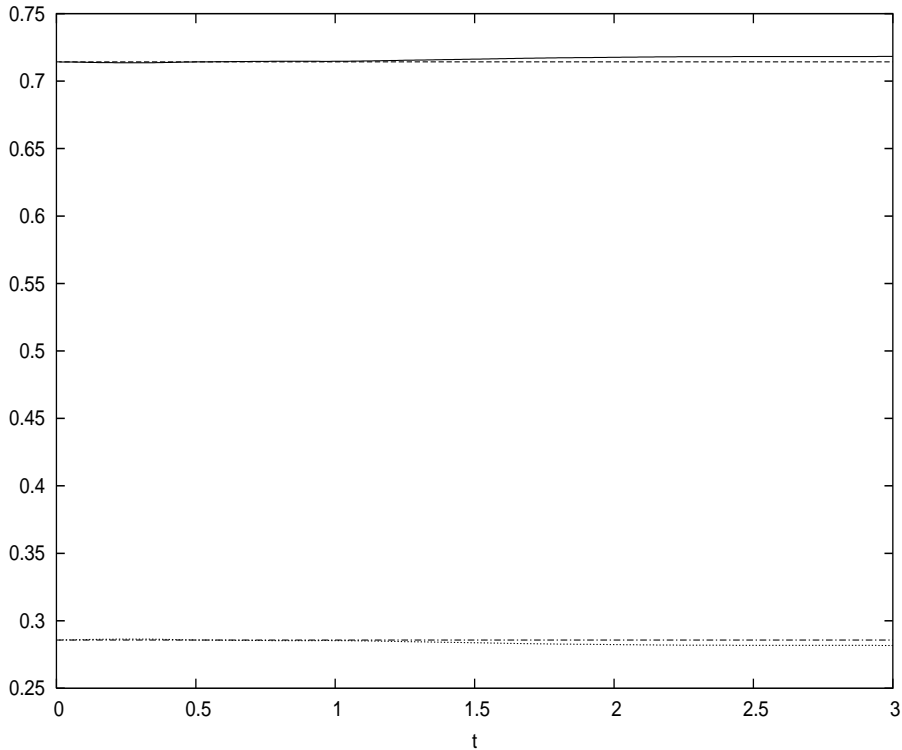,width=5in,height=4in}
\end{figure}

\subsubsection{Example IIIB}

Now consider $H=(\omega/2)\sigma_z$ and $L=(\lambda/2)(\sigma_x-i\sigma_y)$ and again pick $\alpha(t,s)=(\gamma/2)e^{-\gamma(t-s)}$ with $\omega=1$, $\lambda^2=2\omega$ and $\gamma=\omega$. For the initial condition $|\psi_0\rangle =\frac{1}{\sqrt{2}}(|1\rangle +|2\rangle)$ we calculated $\langle 1|\rho_t|2\rangle$ using Eqs. (\ref{lindc}) and an average over 10000 
trajectories. The real and imaginary parts (solid and dotted) are plotted in Fig. 10 against exact results (dashed and dot-dashed). Agreement is good.
\begin{figure}[htp]
\caption{$\langle 1|\rho_t|2\rangle$ for Example IIIB}
\epsfig{file=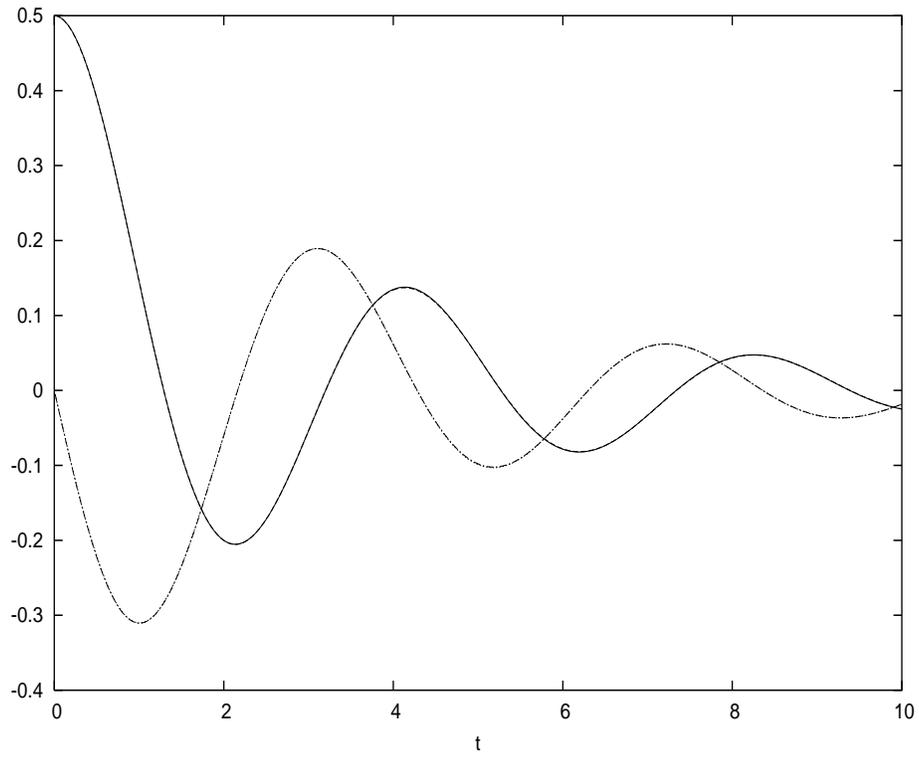,width=5in,height=4in}
\end{figure}
In Fig. 11 we show the numerically computed diagonal matrix elements $\langle 1|\rho_t|1\rangle$ and $\langle 2|\rho_t|2\rangle$ (solid and dotted) against exact results (dashed and dot-dashed). Convergence is again good.
\begin{figure}[htp]
\caption{$\langle 1|\rho_t|1\rangle$ and $\langle 2|\rho_t|2\rangle$ for Example IIIB}
\epsfig{file=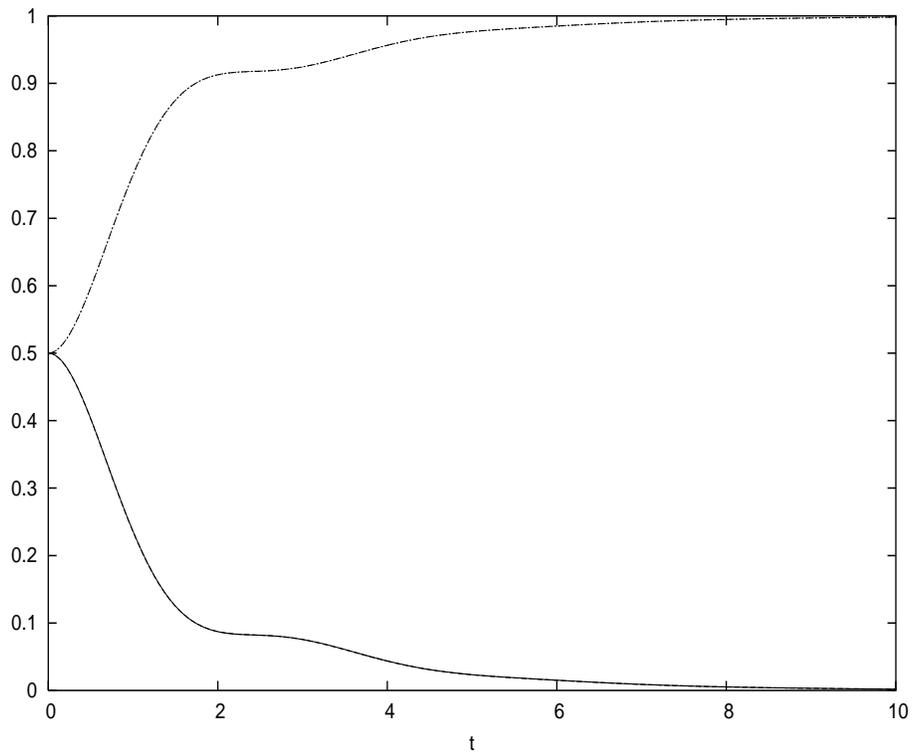,width=5in,height=4in}
\end{figure}

\subsubsection{Example IIIC}

Now consider a harmonic oscillator subsystem with $H=\omega a^{\dag}a$ and $L=\lambda a$. We chose an initial condition $|\psi_0\rangle =\frac{1+2i}{\sqrt{7}}|1\rangle +\frac{1+i}{\sqrt{7}}|2\rangle $  and picked $\alpha(t,s)=(\gamma/2)e^{-\gamma(t-s)}$ with $\omega=1$, $\lambda=\omega$ and $\gamma=\omega$. Convergence is rapid and so we used only 1000 trajectories. The cpu time was about 12 minutes on a 600 MHz Alpha processor. In Fig. 12 we plot the real and imaginary parts (solid and dotted) of $\langle 1|\rho_t|2\rangle$ vs their corresponding exact real and imaginary parts (dashed and dot-dashed). In Fig. 13 we show the diagonal matrix elements $\langle 1|\rho_t|1\rangle$ and $\langle 2|\rho_t|2\rangle$ (solid and dotted) plotted against their corresponding exact results (dashed and dot-dashed). The results are good.
\begin{figure}[htp]
\caption{$\langle 1|\rho_t|2\rangle$ for Example IIIC}
\epsfig{file=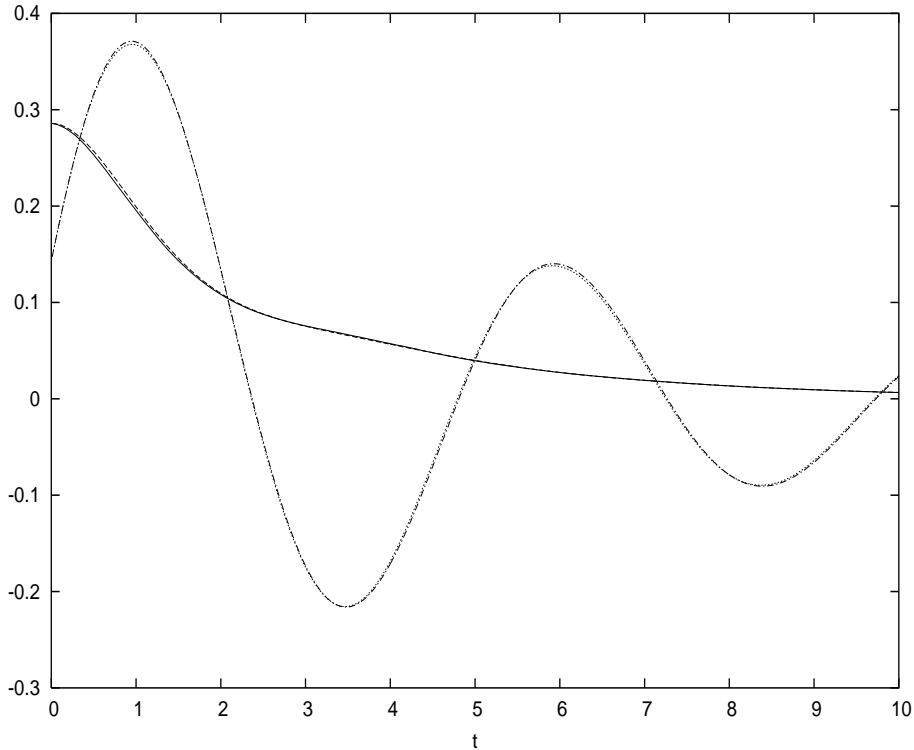,width=5in,height=4in}
\end{figure}
\begin{figure}[htp]
\caption{$\langle 1|\rho_t|1\rangle$ and $\langle 2|\rho_t|2\rangle$ for Example IIIC}
\epsfig{file=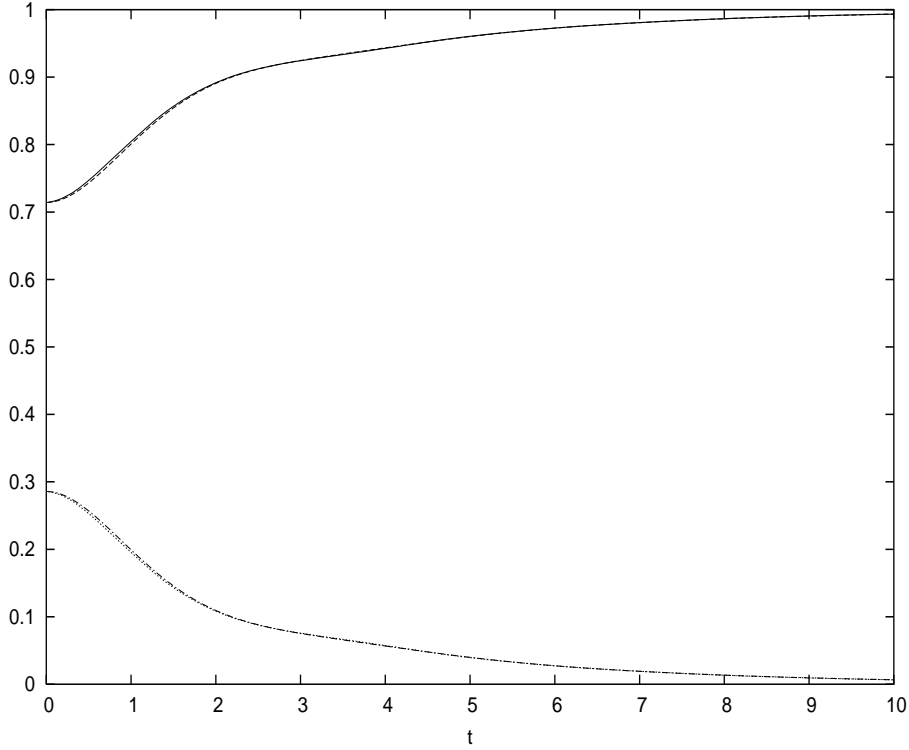,width=5in,height=4in}
\end{figure}

\subsection{General memory function}

When the memory function cannot be represented as a sum of exponentials of the previous form, it is again possible to rewrite the equations in partial differential form. We again introduce a new variable $u$ and define
\begin{equation}
V_{t,u}=\int_0^t ds~\alpha(t+u,s) U_s^{-1}LU_s
\end{equation}
such that the new set of equations has partial differential form
\begin{eqnarray}
\frac{dU_t}{dt}&=&-iH U_t +(z_t+y_{t,0}) ~(L-\langle L\rangle_t) ~U_t \nonumber \\
&-&(L^{\dag}-\langle L^{\dag}\rangle_t)~U_t~V_{t,0} \nonumber \\
&+&
\langle \psi_0|U_t^{\dag}(L^{\dag}-\langle L^{\dag}\rangle_t)U_tV_{t,0} |\psi_0\rangle ~U_t\nonumber \\
\frac{d U_t^{-1}}{dt}&=&iU_t^{-1}H-(z_t +y_{t,0})~U_t^{-1}~(L-\langle L\rangle_t) \nonumber \\
&+&U_t^{-1} (L^{\dag}-\langle L^{\dag}\rangle_t) ~U_t ~V_{t,0}  ~U_t^{-1}\nonumber \\
&-&\langle \psi_0|U_t^{\dag}(L^{\dag}-\langle L^{\dag}\rangle_t)U_tV_{t,0}  |\psi_0\rangle ~U_t^{-1}\nonumber \\
\frac{d V_{t,u}}{dt}&=&\alpha(t+u,t)U_t^{-1}LU_t+\frac{\partial V_{t,u}}{\partial u}\nonumber \\
\frac{dy_{t,u}}{dt}&=&\alpha^*(t+u,t)\langle L^{\dag}\rangle_t+\frac{\partial y_{t,u}}{\partial u}
\end{eqnarray}
where 
\begin{equation}
y_{t,u}=\int_0^t ds~\alpha^*(t+u,s) \langle L^{\dag}\rangle_s.
\end{equation}
The same considerations regarding adding a damping factor and generation of the noise $z_t$, discussed in section IIB, apply here without modification. 

\section{New Examples}

We now consider three new problems which it was previously impossible to explore exactly using NMQSD. 
Our treatment of these
examples is not meant to be exhaustive. We simply wish to show that interesting issues can be explored using
the computational implementation of NMQSD introduced in sections II and III. We choose to use
the norm-preserving version of the theory. This improves convergence and also allows us to study individual
trajectories, which may have some physical significance\cite{Wise}. Specifically, it has been shown that
NMQSD can be interpreted as a hidden variable theory in which the noise $z_t$ represents a hidden-variable of the bath\cite{Wise}. We shall see that individual trajectories do indeed behave in
ways consistent with physical intuition. Whether their non-quantum-mechanical statistical properties 
(e.g. $M[|\langle \psi_0|\psi_t\rangle|^2]=\langle \psi_0|\rho_t|\psi_0\rangle$ can be predicted with standard quantum theory, while $M[|\langle \psi_0|\psi_t\rangle|^4]-M[|\langle \psi_0|\psi_t\rangle|^2]^2$ cannot) are
in agreement with experiment remains an open and interesting question.

\subsection{Example IVA}

Here we consider the dynamics of a symmetric double well representing a reaction coordinate of an isomeric or chiral molecule, interacting with the radiation field. Of course realistic chemical environments contain sources of decoherence other than the radiation field, but the example is still of interest. When an ensemble of such systems is prepared in an initial achiral state, interaction with the radiation field is expected to drive the population to a symmetric final distribution. Indeed, clocks for dating amino acids have been proposed on this basis\cite{Bada}. When the barrier height is low (e.g. in NHDT) individual molecules are observed in states which are superpositions of left and right handed states (e.g. the ground state of the double well). When the barrier height is large, individual molecules are found in left handed states or right handed states but not normally in superpositions (although superpositions can in principle be prepared\cite{Cina}). This is unusual because the eigenstates of the double well are superpositions of left and right handed states. Environment induced superselection rules are sometimes invoked to explain this effect\cite{Dec,Math}. We will now explore the predictions of NMQSD in these two cases.

Consider a quartic oscillator subsystem with $H=\omega [a^{\dag}a-(3/8)(a^{\dag}+a)^2+\epsilon(a^{\dag}+a)^4]$ and $L=\lambda a$. This Hamiltonian corresponds to a symmetric double well potential and so could represent a reaction coordinate for isomerization or racemization of a molecule. We choose an initial condition $|\psi_0\rangle =\frac{1}{\sqrt{2}}(|0\rangle -|1\rangle) $  which means the particle is initially localized in the left well, and pick $\alpha(t,s)=(\gamma/2)e^{-\gamma(t-s)}$ with $\omega=1$, $\epsilon=\hbar\omega/E_b$, $\lambda=\omega$ and $\gamma=\omega$ where $E_b$ is the activation energy of the barrier. The parameter $\epsilon$ controls the effective barrier height.

First consider the case where $\epsilon=.692$, which is typical for proton transfer isomerization\cite{MM}. In Fig. 14 we plot the probability density $\langle x|\rho_t|x\rangle$ against $x$ (in units of $\sqrt{\hbar/m\omega}$) for times $t=0$ (solid), 2 (dashed), 4 (dotted), 6 (dot-dashed) and 12 (double-dashed). Dissipation and decoherence drive the population from a nearly pure left-handed state to a symmetric mixture. The calculation was performed in a basis set of the lowest five harmonic oscillator states. A total of 1000 trajectories were included in the average. The calculation again took about ten minutes.
\begin{figure}[htp]
\caption{Relaxation of $\langle x|\rho_t|x\rangle$ vs $x$ for Example IVA with $\epsilon=.692$}
\epsfig{file=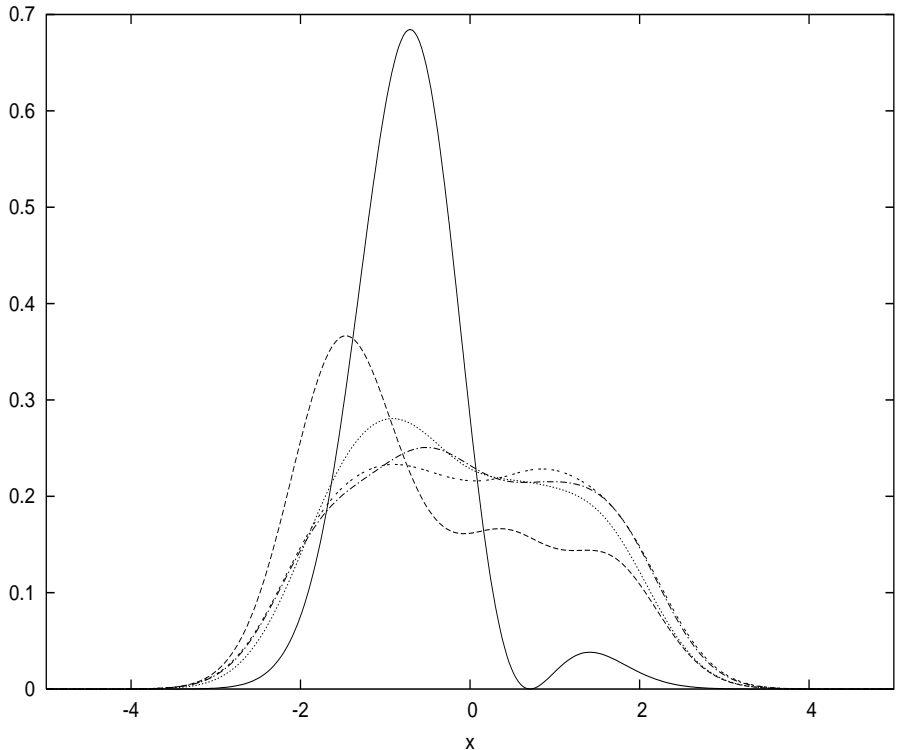,width=5in,height=4in}
\end{figure}
In Fig. 15 we plot $\langle x|\rho_t|x\rangle$ against $x$, computed for 1000 and 10000 trajectories, at times $t=2$ (solid and dashed), 4 (dotted and dot-dashed) and 12 (double-dashed and triple-dashed). Convergence is again quite good for 1000 trajectories.
\begin{figure}[htp]
\caption{Convergence of $\langle x|\rho_t|x\rangle$ vs $x$ for Example IVA with $\epsilon=.692$}
\epsfig{file=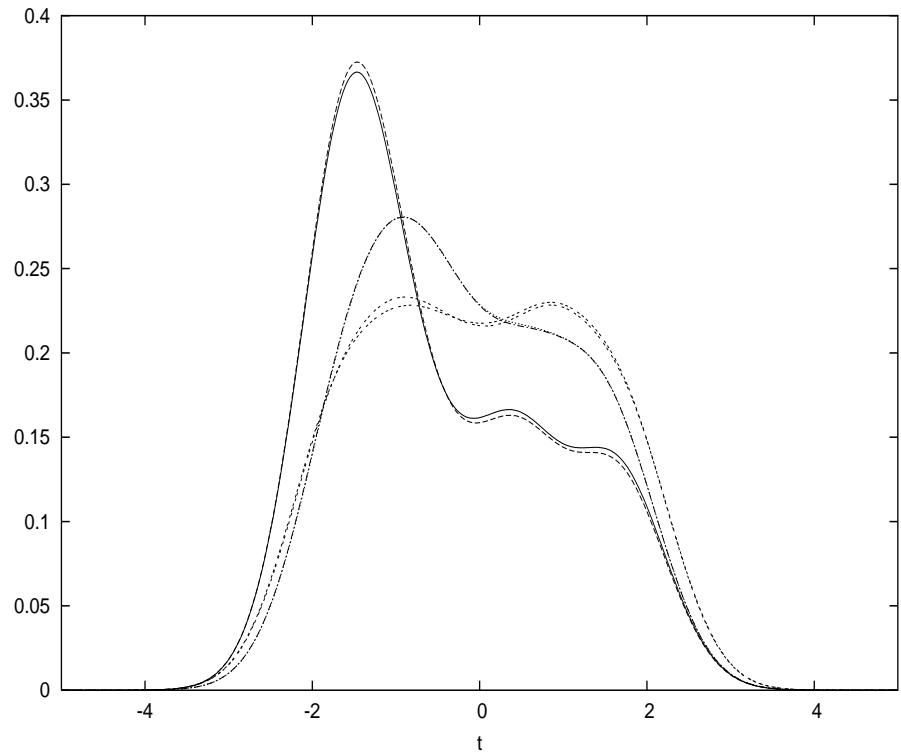,width=5in,height=4in}
\end{figure}

Assuming that individual trajectories may have some physical significance, as has been suggested\cite{Wise}, it is worth examining a few to see whether their dynamics makes intuitive sense. For this moderate barrier case we anticipate asymptotic states which are mixtures of left and right. In fact as we see in Fig. 16, where we plot densities for individual trajectories at time $t=12$,
\begin{figure}[htp]
\caption{$\langle x|\psi_t\rangle\langle\psi_t|x\rangle$ vs $x$ for individual trajectories at $t=12$, for Example IVA with $\epsilon=.692$ }
\epsfig{file=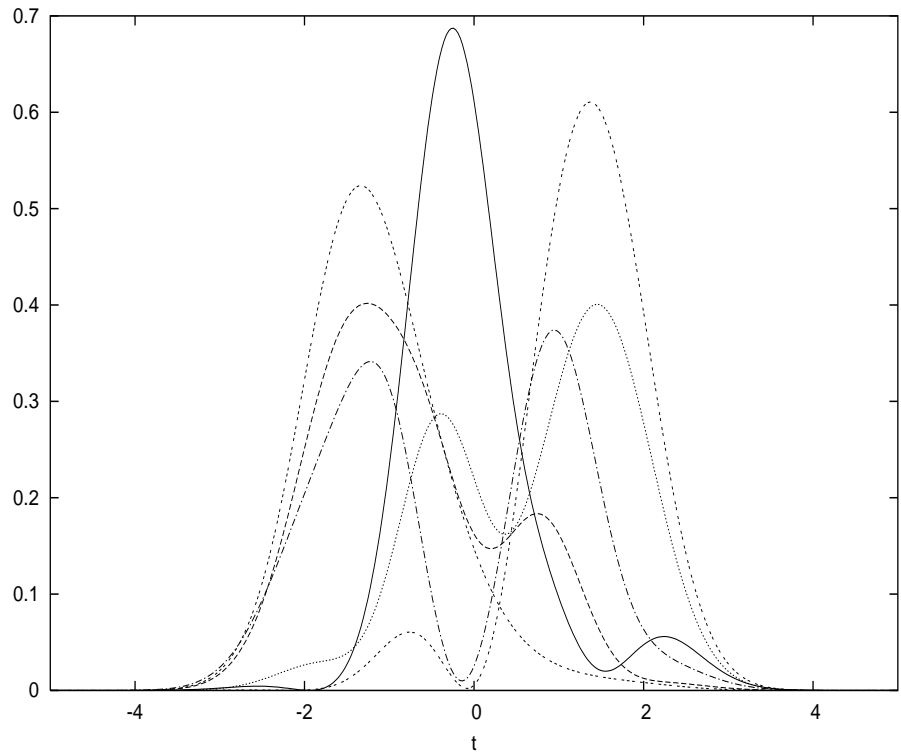,width=5in,height=4in}
\end{figure}
some trajectories do lead to superpositions. However, some also remain strongly localized on the left of the barrier, while others have made a jump to the right. Thus, superselection appears to begin to play a role even for moderate barriers.

Now consider the case of a higher barrier, where $\epsilon=.1$, which is still very modest compared to that expected for large chiral molecules. In Fig. 17 we again plot $\langle x|\psi_t\rangle\langle\psi_t|x\rangle$ vs $x$ for individual trajectories, this time at $t=14$.
\begin{figure}[htp]
\caption{$\langle x|\psi_t\rangle\langle\psi_t|x\rangle$ vs $x$ for individual trajectories at $t=14$, for Example IVA with $\epsilon=.1$}
\epsfig{file=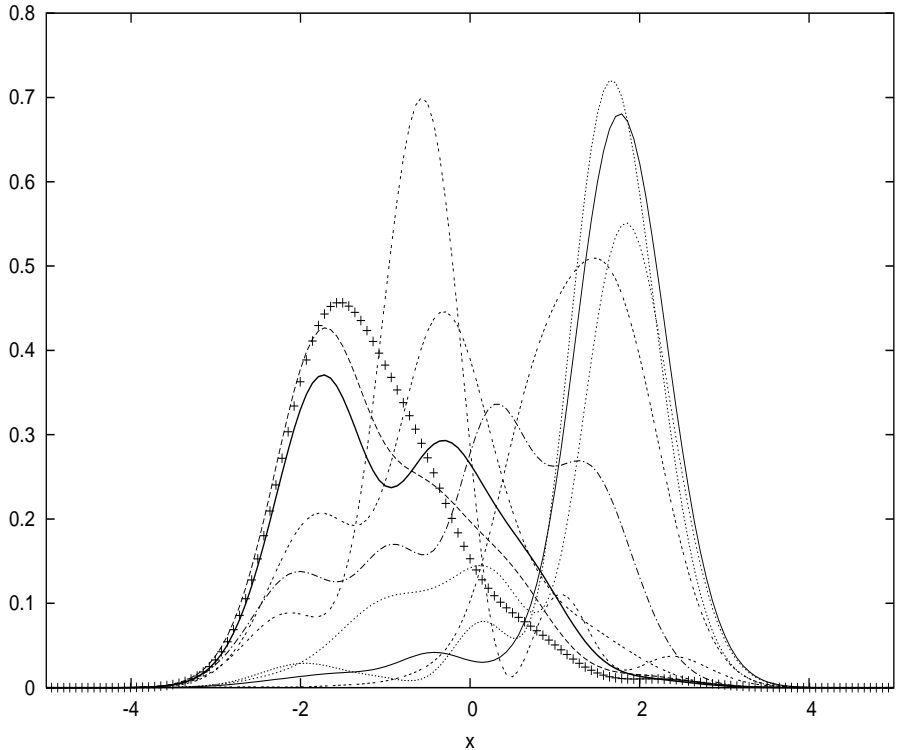,width=5in,height=4in}
\end{figure}
For this higher barrier the individual densities are more strongly chiral, with most strongly localized on one side of the barrier or the other. The only state even close to an equal superposition of left and right is the dot-dashed curve. These results support the notion of the emergence of a superselection rule favoring states localized on one side of the barrier or the other over superpositions.

In Fig. 18 we follow a single trajectory as it tunnels from one side of the barrier to the other.  
\begin{figure}[htp]
\caption{$\langle x|\psi_t\rangle\langle\psi_t|x\rangle$ vs $x$ for an individual trajectory at various times, for Example IVA with $\epsilon=.1$}
\epsfig{file=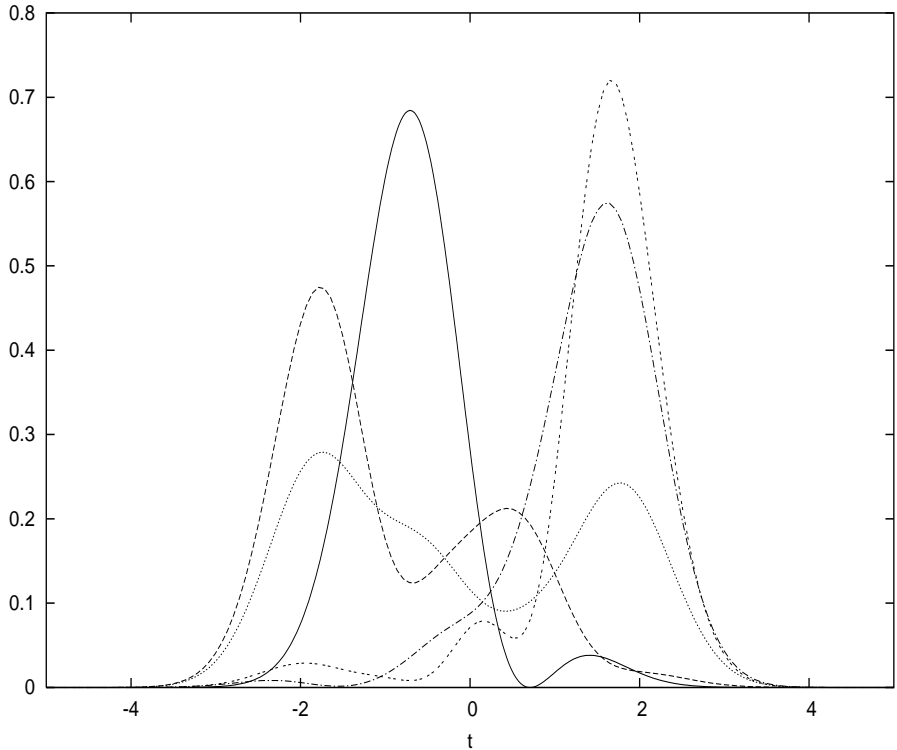,width=5in,height=4in}
\end{figure}
The densities indicated are for times $t=0$ (solid), 2 (dashed), 4 (dotted), 6 (dot-dashed) and 14 (double-dashed). The tunneling process appears to start almost immediately and is effectively over by $t=6$. In other cases tunneling had already occurred by $t=2$. This apparent discrepancy invites further scrutiny in light of the many
studies predicting well defined tunneling times\cite{TT}.

\subsection{Example IVB}

Here we consider a two-level atom immersed in the radiation field of a dielectric band gap. This model has
recently been studied using approximations to the NMQSD equations\cite{Gasp}. In an interaction picture,
rotating with the subsystem Hamiltonian, $H=0$ and $L=\frac{i\lambda}{2}(\sigma_x-i\sigma_y)$. The memory
function is given by
\begin{equation}
\alpha(t,s)=e^{i(\omega-A)(t-s)}[J_0(\frac{B}{3}(t-s))]^3\label{MEM4}
\end{equation}
where $\hbar\omega$ is the excitation energy of the atom, a band of allowed frequencies lies between $A-B$ and $A+B$, the gap lies between 0 and $A-B$, and $J_0$ is the Bessel function of the first kind of order 0. We scale time in units of $\hbar/\lambda$ and energy in units of $\lambda$. In these units we set $\omega=10$, $B=5$ and we will consider two values of $A$ corresponding to $\omega$ in the band ($A=\omega$) and in the gap ($A=\omega+B+1$).
\begin{center}
\begin{tabular}{|c|c|c|}
\hline
$A_j$ & $\gamma_j$ & $\pm \omega_j^0$ \\ \hline
  6.636$\times 10^{-2}$  & 3.3602$\times 10^{-3}$  & 0.9598   \\
  1.597$\times 10^{-2}$  & 0.2301 &       4.4870     \\
  6.513$\times 10^{-2}$  & 1.0000$\times 10^{-5}$ & 0.5762   \\
  2.112$\times 10^{-2}$  & 0.2003 &       4.0230    \\
  6.551$\times 10^{-2}$  & 1.0000$\times 10^{-5}$ &  0.1924    \\
  2.202$\times 10^{-2}$  & 0.1335  &    3.6070   \\
  3.460$\times 10^{-2}$  & 2.1033$\times 10^{-2}$ &  2.4534   \\
  4.018$\times 10^{-2}$  & 2.1043$\times 10^{-4}$ &  2.0778    \\
  2.619$\times 10^{-2}$  & 8.8754$\times 10^{-2}$ &   3.2106    \\
  5.637 $\times 10^{-2}$ & 1.0000$\times 10^{-5}$ &   1.7026   \\
  2.731 $\times 10^{-2}$ & 3.8032 $\times 10^{-2}$ &  2.8303    \\
  6.388 $\times 10^{-2}$ & 1.0000$\times 10^{-5}$ &   1.3391  \\
\hline
\end{tabular}
\end{center}

We fitted (\ref{MEM4}) to the form (\ref{MEM}) using nonlinear least squares, where $\omega_j$ come in pairs $\omega_j=\omega-A+\omega_j^0$ and $\omega_j=\omega-A-\omega_j^0$ for each $A_j$ and $\gamma_j$. The resulting parameters are given in Table 1. In Fig. 19 we plot (\ref{MEM4}) (solid curve) against (\ref{MEM}) (dashed) , which shows that the fit is quite satisfactory.
\begin{figure}[htp]
\caption{$\alpha(t,0)$ for Example IVB with $\omega=A$}
\epsfig{file=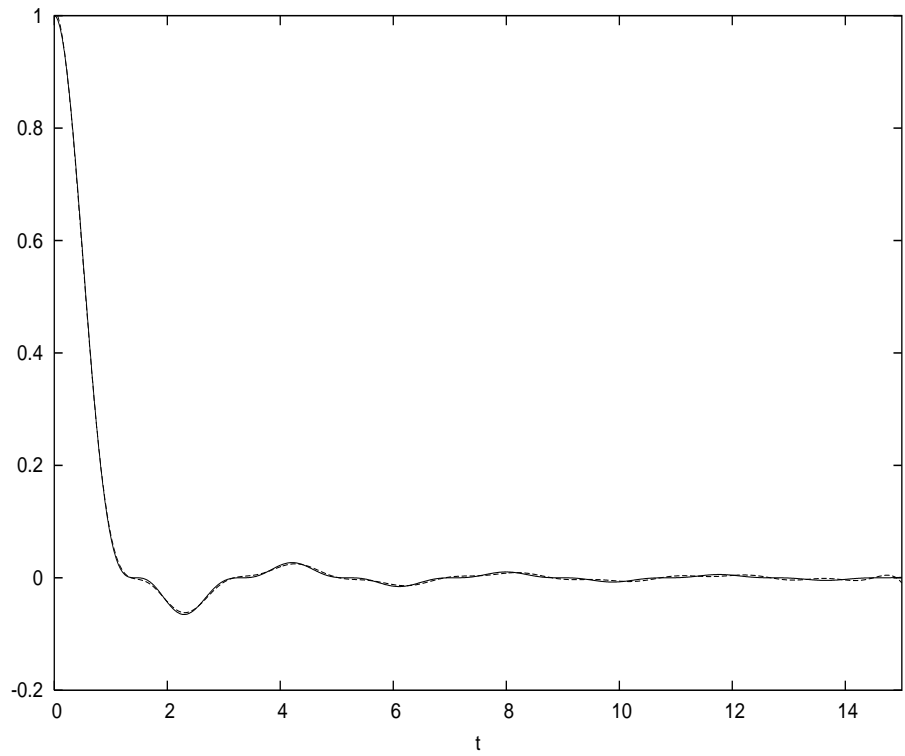,width=5in,height=4in}
\end{figure}

We chose an initial state $|\psi_0\rangle=|1\rangle$ which corresponds to an excited 2-level atom (note that our
convention for labeling the states is the reverse of that in Ref. \cite{Gasp}). We calculated $\langle 1|\rho_t|1\rangle$ and $\langle 2|\rho_t|2\rangle$ using 1000 trajectories for $\omega$ in the band ($\omega=A$, solid and dotted curves) and $\omega$ in the gap ($\omega=A+B+1$, dashed and dot-dashed curves). These quantities are plotted in Fig. 20. When $\omega$ is in the band, emission of a photon occurs, and the two-level atom relaxes to its ground state. When $\omega$ is in the gap the system evolves toward a statistical superposition of the two states (i.e. $\langle 1|\rho_t|2\rangle=0$) which is weighted toward the excited state. 
\begin{figure}[htp]
\caption{$\langle 1|\rho_t|1\rangle$ and $\langle 2|\rho_t|2\rangle$ for Example IVB with $\omega=A$ ( solid and dotted) and $\omega=A+1$ (dashed and dot-dashed)}
\epsfig{file=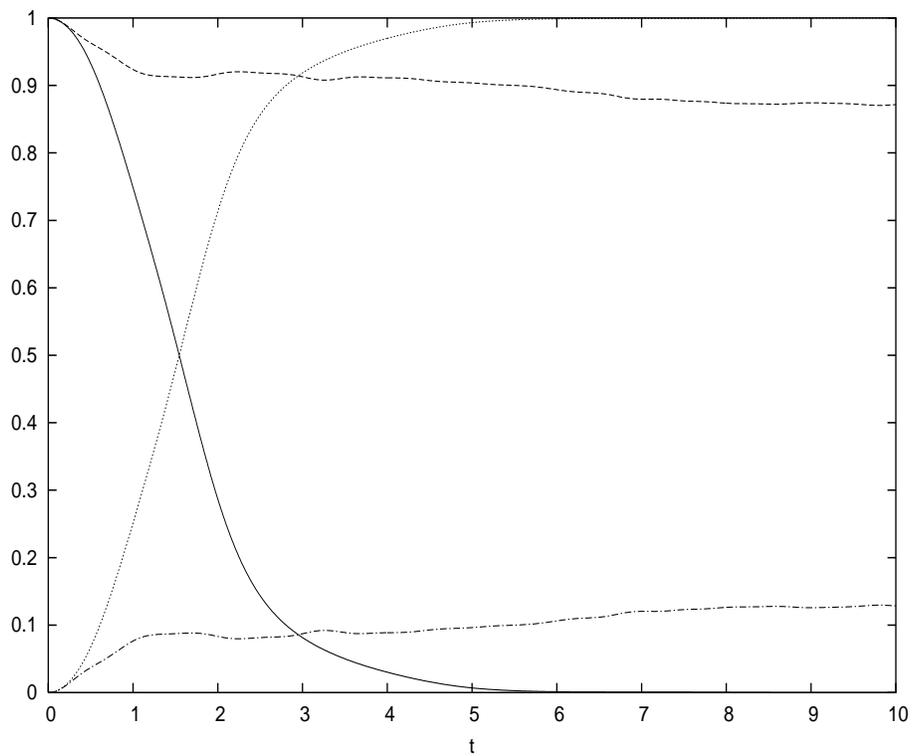,width=5in,height=4in}
\end{figure}
It should however be remembered that each individual trajectory represents a pure state and so has a density matrix with nonzero off-diagonal elements. In Fig. 21 we show the density matrix elements $\langle 1|\rho_t|1\rangle$ (solid) and $\langle 2|\rho_t|2\rangle$ (dashed) and the real and imaginary parts of $\langle 1|\rho_t|2\rangle$ (dotted and dot-dashed) for an individual trajectory with $\omega$ in the gap.
\begin{figure}[htp]
\caption{$\langle 1|\rho_t|1\rangle$, $\langle 2|\rho_t|2\rangle$ and $\langle 1|\rho_t|2\rangle$ for a single trajectory of Example IVB}
\epsfig{file=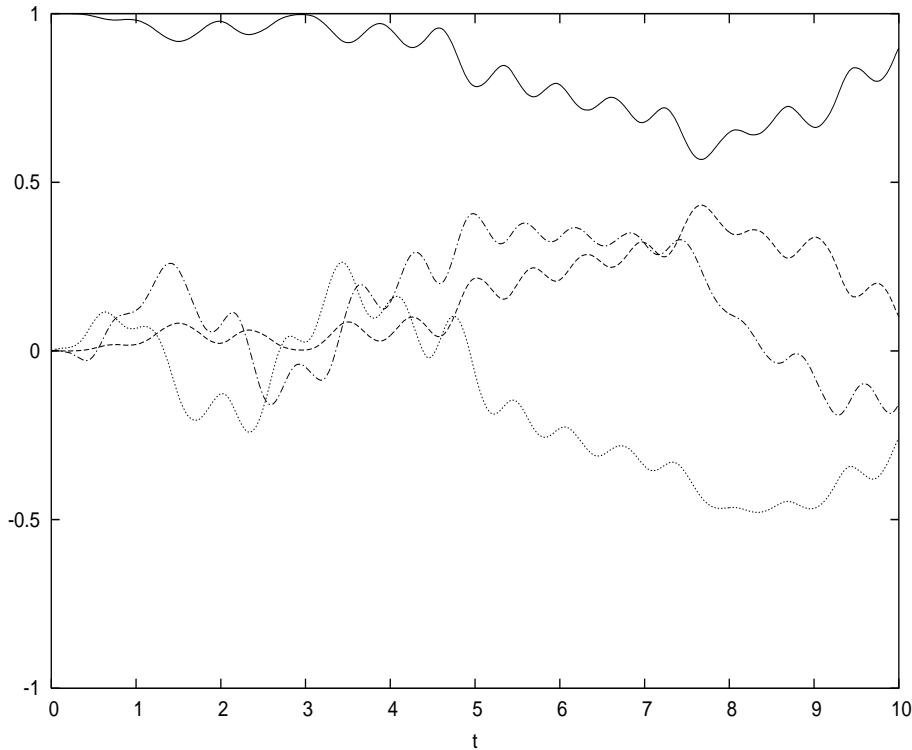,width=5in,height=4in}
\end{figure}
Individual trajectories thus evolve toward a {\em coherent} superposition of the excited and ground states, with the excited state weighted more heavily.

\subsection{Example IVC}

Finally, we consider a three-level system ($|1\rangle$ with energy $E_1=0$ in eV, $|2\rangle$ with $E_2=4.4$, and $|3\rangle$ with $E_3=0$) representing three electronic states of a Mg$^+$ ion in an ion trap, which is driven by a laser resonant with the transition between levels 1 and 2. This system has been quite extensively studied\cite{Jump,Master,Master2,Knight,JumpTh} due to the interesting phenomenon of intermittent fluorescence exhibited by the ion. The ion which normally cycles between states 1 and 2 occasionally jumps into the dark state 3 giving rise to periods where no fluorescence is observed. These stochastic jumps have been observed experimentally\cite{Jump} and theoretically\cite{Master,JumpTh} over a time scale with millisecond resolution. It has been suggested that such stochastic jumps might occur on all time scales\cite{JumpTh}. Here we show that NMQSD predicts jump phenomena on a picosecond timescale consistent with this scenario.

The effective Hamiltonian in a frame rotating with the Hamiltonian of the isolated ion has the form
\begin{equation}
H=(\Omega/2)(|1\rangle \langle 2|+|2\rangle \langle 1|)
\end{equation}
where $\Omega$ is the Rabi frequency of the 1-2 transition. The coupling operators $L_i$ for $i=1,\dots, 4$ have the forms
\begin{eqnarray}
L_1&=&\lambda_{12}|1\rangle \langle 2|\nonumber \\
L_2&=&\lambda_{13}|1\rangle \langle 3|\nonumber \\
L_3&=&\lambda_{31}|3\rangle \langle 1|\nonumber \\
L_4&=&\lambda (|1\rangle\langle 1|-|3\rangle\langle 3|).
\end{eqnarray}
The parameters were chosen as $\lambda_{12}=\sqrt{\gamma/\tau}$, $\lambda_{13}=\sqrt{R_-/\tau}$, $\lambda_{31}=\sqrt{R_+/\tau}$ where $\gamma$ is the spontaneous decay rate for the 2-1 transition and $R_-$ and $R_+$ are the rates out of and into the dark state 3, respectively\cite{Master2}. It has been shown that
$R_-=8\Omega^2\gamma/9\alpha^2$ and $R_+=\Omega^2\gamma/18\alpha^2$ where $\alpha$ denotes a Zeeman splitting\cite{Master2}. Finally, the $\lambda$ operator arises due to interaction with the photodetectors. To reproduce the results of Ref. \cite{Master2} in the Markovian limit we set $\tau=\int_0^{\infty} {\rm Re} ~\alpha (t) dt$ . The memory function, which is common to the four noises, was calculated with a Debye distribution of frequencies and for a temperature of about .1 K. We chose time units $10^{-3}\gamma^{-1}$ - roughly a picosecond since $\gamma=2\pi 43$ MHz\cite{Master2} - in terms of which we set $\alpha=12.1$, $\Omega=2$, $\lambda_{12}=10^{-3}/\sqrt{\tau}$, $\lambda_{31}=\sqrt{1/18}\Omega/(\alpha \sqrt{\tau})$, $\lambda_{13}=\sqrt{8/9}\Omega/(\alpha \sqrt{\tau})$ and $\lambda=.22/sqrt{\tau}$. We calculated that $\tau=509$ from the 5 term non-linear least squares fit to the memory function, although the principle decay occurs over the first 50 time units. The dynamics is thus quite strongly non-Markovian.
\begin{figure}[htp]
\caption{$|\langle 1|\psi_t\rangle |^2+|\langle 2|\psi_t\rangle |^2$ vs $t$ for a single trajectory of Example IVC}
\epsfig{file=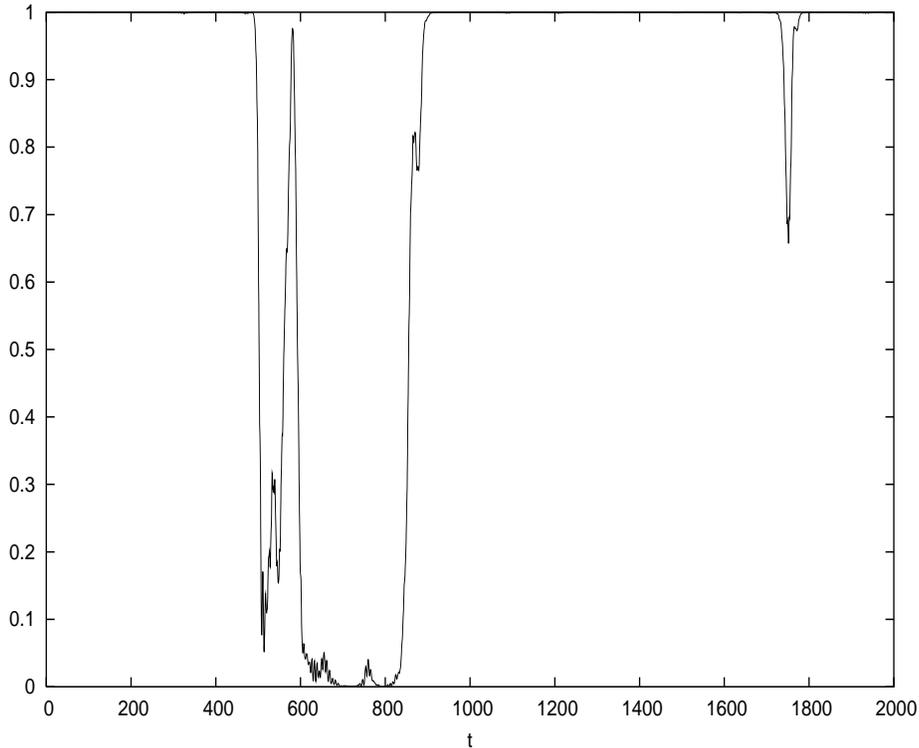,width=5in,height=4in}
\end{figure}

In Fig. 22 we show the occupation probability of the 1-2 manifold $|\langle 1|\psi_t\rangle |^2+|\langle 2|\psi_t\rangle |^2$ plotted against time. At various times and for various lengths of time this probability approaches zero indicating that the ion has jumped to the dark state 3. A more detailed analysis of the statistical properties of the dark periods will be presented elsewhere.

\section{Discussion}

Non-Markovian quantum state diffusion (NMQSD), being an exact unraveling of the master equation for an arbitrary subsystem interacting linearly with a boson bath, potentially has a wide range of applications in quantum optics.
This potential has not been realized to any significant extent due to the difficulty of solving - even numerically - the variational-differential evolution equation. In fact the development of NMQSD has not led to the 
exact solution of any new problems. In this manuscript we have shown how the variational-differential equations (VDEs) of NMQSD can be exactly rewritten as integrodifferential equations which can in turn be rewritten as ordinary or partial differential equations. We illustrated application of the new NMQSD equations by solving a number of problems for which exact solutions were already known. Both linear and nonlinear versions of NMQSD were 
studied. We found that both versions worked well and yielded high accuracy solutions. Finally, we applied the method to three previously unsolvable problems (tunneling in a double well, two-level atom in a photonic band gap, and intermittent fluorescence in a driven three-level ion) to show that interesting work 
can be done with the reformulated theory. We anticipate that other interesting applications of NMQSD will emerge now that the equations can be solved in a systematic fashion.

The authors acknowledge the support of the Natural Sciences and 
Engineering Research Council of Canada.


\begin{thebibliography}{99}
\bibitem{Cnoise} We use the simple notation $z_t$ rather than the physically suggestive $z_t^*$\cite{Wise}.
\end{thebibliography}
\end{document}